\documentclass[10pt,journal,compsoc]{IEEEtran}

\usepackage{multirow}
\usepackage{amsmath,amssymb,amsfonts}
\usepackage{algorithmic}
\usepackage{graphicx}
\usepackage{textcomp}
\usepackage{xcolor}
\usepackage{adjustbox}
\usepackage{multirow}

\usepackage{tikz}
\usepackage{pgfmath}

\usepackage{algorithm2e}
\usepackage{booktabs}
\usepackage{caption}

\usetikzlibrary{arrows.meta, graphs, shapes.misc}
\usetikzlibrary{datavisualization}
\usetikzlibrary{circuits.logic.US}

\definecolor{cayenne}{HTML}{791C22}
\definecolor{carrot}{HTML}{007BA7}
\definecolor{banana}{HTML}{FFE135}
\definecolor{salmon}{HTML}{FBCEB1} %F4AE72}
\definecolor{amber}{HTML}{FF7E00} %F4AE72}
\definecolor{fusia}{HTML}{915C83} %F4AE72}
\definecolor{yllow}{HTML}{E9D66B} %F4AE72}
\definecolor{scarl}{HTML}{8FBC8F} %F4AE72}
\definecolor{honey}{HTML}{D4AF37} %E5AA70
\definecolor{ash}{HTML}{EFDECD} %A1CAF1
\definecolor{bitter}{HTML}{8FBC8F}
\definecolor{celadon}{HTML}{ED9121}

\definecolor{vred}{HTML}{E5B9B5}
\definecolor{vgreen}{HTML}{C4D6A0}

\usetikzlibrary{positioning}

\def\squarecorner#1{

    \pgf@x=\the\wd\pgfnodeparttextbox%
    \pgfmathsetlength\pgf@xc{\pgfkeysvalueof{/pgf/inner xsep}}%
    \advance\pgf@x by 2\pgf@xc%
    \pgfmathsetlength\pgf@xb{\pgfkeysvalueof{/pgf/minimum width}}%
    \ifdim\pgf@x<\pgf@xb%
        \pgf@x=\pgf@xb%
    \fi%

    \pgf@y=\ht\pgfnodeparttextbox%
    \advance\pgf@y by\dp\pgfnodeparttextbox%
    \pgfmathsetlength\pgf@yc{\pgfkeysvalueof{/pgf/inner ysep}}%
    \advance\pgf@y by 2\pgf@yc%
    \pgfmathsetlength\pgf@yb{\pgfkeysvalueof{/pgf/minimum height}}%
    \ifdim\pgf@y<\pgf@yb%
        \pgf@y=\pgf@yb%
    \fi%

    \ifdim\pgf@x<\pgf@y%
        \pgf@x=\pgf@y%
    \else
        \pgf@y=\pgf@x%
    \fi

    \pgf@x=#1.5\pgf@x%
    \advance\pgf@x by.5\wd\pgfnodeparttextbox%
    \pgfmathsetlength\pgf@xa{\pgfkeysvalueof{/pgf/outer xsep}}%
    \advance\pgf@x by#1\pgf@xa%

    \pgf@y=#1.5\pgf@y%
    \advance\pgf@y by-.5\dp\pgfnodeparttextbox%
    \advance\pgf@y by.5\ht\pgfnodeparttextbox%
    \pgfmathsetlength\pgf@ya{\pgfkeysvalueof{/pgf/outer ysep}}%
    \advance\pgf@y by#1\pgf@ya%
}
\makeatother

\pgfdeclareshape{square}{
    \savedanchor\northeast{\squarecorner{}}
    \savedanchor\southwest{\squarecorner{-}}

    \foreach \x in {east,west} \foreach \y in {north,mid,base,south} {
        \inheritanchor[from=rectangle]{\y\space\x}
    }
    \foreach \x in {east,west,north,mid,base,south,center,text} {
        \inheritanchor[from=rectangle]{\x}
    }
    \inheritanchorborder[from=rectangle]
    \inheritbackgroundpath[from=rectangle]
}

\ifCLASSOPTIONcompsoc
  \usepackage[nocompress]{cite}
\else
  \usepackage{cite}
\fi

\begin{document}

\title{Near-Precise Parameter Approximation for Multiple Multiplications on A Single DSP Block}

\author{Ercan~Kalali
        and Rene~van~Leuken

\IEEEcompsocitemizethanks{\IEEEcompsocthanksitem E. Kalali and R. van Leuken are with the Department
of Microelectronics, Delft University of Technology, Delft,
2628 CD, The Netherlands.\protect\\
E-mail: e.kalali-1@tudelft.nl, t.g.r.m.vanleuken@tudelft.nl}% <-this % stops a space
}

\IEEEtitleabstractindextext{%
\begin{abstract}
A multiply-accumulate (MAC) operation is the main computation unit for DSP applications. DSP blocks are one of the efficient solutions to implement MACs in FPGA's. However, since the DSP blocks have wide multiplier and adder blocks, MAC operations using low bit-length parameters lead to an underutilization problem. Hence, an efficient approximation technique is introduced. The technique includes manipulation and approximation of the low bit-length fixed-point parameters based upon a Single DSP - Multiple Multiplication (SDMM) execution. The SDMM changes the traditional MAC implementation in the DSP block by separating multiplication and accumulation operations. While the accumulator hardware available in the DSP block is used for multiple parameter multiplication, parallel LUTs are employed for the accumulation part of the MAC operation. The accuracy of the developed optimization technique was evaluated for different CNN weight bit precisions using the Alexnet and VGG-16 networks and the Tiny ImageNet dataset. The optimization can be implemented without loss of accuracy in almost all cases, while it causes slight accuracy losses in a few cases. Through these optimizations, the SDMM is performed at the cost of a small hardware overhead. For example, a single DSP block executes 3 8-bit fixed-point parameter multiplications. As a result of our optimizations, the parameters are represented in a different format on off-chip memory, providing up to 33\% compression without any hardware cost. The compression rate can be further increased by up to 97\% when used in conjunction with other compression methods for the VGG-16. Reaching this compression rate requires extra hardware cost. A prototype systolic array architecture was implemented employing our optimizations on a Xilinx Zynq FPGA. It reduced the number of DSP blocks by 66.6\%, 75\%, and 83.3\% for 8, 6, and 4-bit input variables, respectively.
\end{abstract}

}

\maketitle
\IEEEdisplaynontitleabstractindextext
\IEEEpeerreviewmaketitle

\ifCLASSOPTIONcompsoc
\IEEEraisesectionheading{\section{Introduction}\label{sec:introduction}}
\else
\section{Introduction}
\label{sec:introduction}
\fi
\IEEEPARstart{T}{he} multiply-accumulate (MAC) operation is the main computation unit for many digital signal processing applications such as convolutional neural networks (CNN), image/video processing, and audio recognition \cite{e1} -\cite{e5}. State-of-the-art CNN algorithms consist of multiple convolutional layers, each requiring millions of MAC operations. This substantial computational complexity has led GPU and FPGA vendors to offer optimized solutions.

DSP blocks in FPGAs are one of the solutions to perform multiply-accumulate (MAC) operations efficiently. These blocks have built-in multiplier and accumulator accelerators instead of using look-up tables (LUTs) to provide high performance and low power. For example, the Xilinx DSP48E1 has a 25$\times$18-bit multiplier, 25-bit pre-adder, and 48-bit accumulator. Each DSP block can perform one MAC operation with high performance and low power consumption. Hence, the number of parallel MAC operations implemented is restricted given the number of available DSP blocks on the FPGA. Furthermore, DSP48 blocks are placed on multiple DSP columns on the FPGA architecture, which enables the implementation of multiple MACs efficiently by cascading multiple DSP48 blocks in each DSP column. The cascaded connection of the DSP blocks is restricted given the number of DSP48 blocks in each DSP column. There are a few studies available in the literature on the efficient use of DSP blocks to overcome these limitations \cite{b4} -\cite{b7}.

We introduce an efficient parameter approximation technique that performs the Single DSP - Multiple Multiplication (SDMM) and increases the DSP utilization ratio. We start with the manipulation described in \cite{c1}. Since the manipulation in \cite{c1} is only employable for a few pre-computed constants, we first modified it to support performing fixed-point signed parameter manipulation during hardware runtime. This modification allows the dynamic control of manipulation of the millions of parameters to use for the SDMM on the FPGA. We achieve this using a table lookup technique. Further, we implemented multiple signed fixed-point parameter multiplications on a single DSP. The traditional MAC implementation in the DSP block is changed by separating multiplication and accumulation operations to implement the SDMM. While the accumulator hardware available in DSP block is used for multiple parameter multiplication, parallel LUTs are employed for the accumulation part of the MAC operation. 

We introduce a novel parameter approximation technique that modifies the parameter's value to guarantee that the bit-length of the manipulated parameter is at most 3, which reduces the overhead caused by the lookup table used to store manipulated parameters significantly. It also makes it easier to multiply more parameters on a single DSP block since the approximation reduces the bit-length of the manipulated parameters. Additionally, it significantly simplifies the control complexity of the hardware implementation. To make sure that all parameter tuples can be implemented using the SDMM, a novel fine-tuning technique is implemented, which ensures the number of parameter multiplication per DSP block is fixed. 

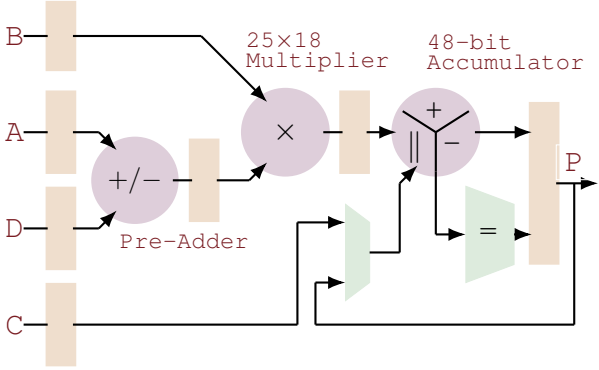
\begin{figure}[htbp]
\centering
\begin{tikzpicture}[INode/.style={circle, draw, square, thick, color=salmon, fill=salmon, minimum size=8}, 
                    PENode/.style={circle, draw, square, thick, color=cayenne, minimum size=13}, scale=0.8]

\node[rectangle, draw, ultra thick, color = ash, fill=ash, minimum size = 10, minimum height = 25] at (0.87,0) {};
\node[draw, color=white] at (0.1,0) {\color{cayenne} \large \texttt{B}};
\draw[thick, color=black] (0.25,0)--(0.65,0);

\node[rectangle, draw, ultra thick, color = ash, fill=ash, minimum size = 10, minimum height = 30] at (0.87,-1.6) {};
\node[draw, color=white] at (0.1,-1.6) {\color{cayenne} \large \texttt{A}};
\draw[thick, color=black] (0.25,-1.6)--(0.65,-1.6);

\node[rectangle, draw, ultra thick, color = ash, fill=ash, minimum size = 10, minimum height = 30] at (0.87,-3.2) {};
\node[draw, color=white] at (0.1,-3.2) {\color{cayenne} \large \texttt{D}};
\draw[thick, color=black] (0.25,-3.2)--(0.65,-3.2);

\node[rectangle, draw, ultra thick, color = ash, fill=ash, minimum size = 10, minimum height = 30] at (0.87,-4.8) {};
\node[draw, color=white] at (0.1,-4.8) {\color{cayenne} \large \texttt{C}};
\draw[thick, color=black] (0.25,-4.8)--(0.65,-4.8);

% pre-adder
\node[circle, draw, ultra thick, color = fusia!30, fill=fusia!30, minimum size = 5, minimum height = 5] at (2.1,-2.4) {\color{black} \texttt{$+/-$}};
\node[text width=2cm] at (3.1,-3.4) {\color{cayenne} \small \texttt{Pre-Adder}};

\draw[thick, color=black, arrows = {-Latex[width=5pt]}] (1.5,-1.6)--(1.8,-1.9);
\draw[thick, color=black] (1.07,-1.6)--(1.5,-1.6);

\draw[thick, color=black, arrows = {-Latex[width=5pt]}] (1.5,-3.2)--(1.8,-2.9);
\draw[thick, color=black] (1.07,-3.2)--(1.5,-3.2);

\node[rectangle, draw, ultra thick, color = ash, fill=ash, minimum size = 10, minimum height = 30] at (3.25,-2.4) {};
\draw[thick, color=black] (2.73,-2.4)--(3.05,-2.4);

% multiplier
\node[circle, draw, ultra thick, color = fusia!30, fill=fusia!30, minimum size = 32, minimum height = 32] at (4.6,-1.6) {\color{black} \large \texttt{$\times$}};
\node[text width=2cm] at (5.2,-0.08) {\color{cayenne} \small \texttt{25$\times$18}};
\node[text width=2cm] at (5.2,-0.42) {\color{cayenne} \small \texttt{Multiplier}};

\draw[thick, color=black, arrows = {-Latex[width=5pt]}] (3.2,0)--(4.3,-1.1);
\draw[thick, color=black] (1.07,0)--(3.2,0);

\draw[thick, color=black, arrows = {-Latex[width=5pt]}] (4.0,-2.4)--(4.3,-2.1);
\draw[thick, color=black] (3.46,-2.4)--(4.0,-2.4);

% accumulator
\node[rectangle, draw, ultra thick, color = ash, fill=ash, minimum size = 10, minimum height = 30] at (5.75,-1.6) {};
\draw[thick, color=black] (5.23,-1.6)--(5.55,-1.6);

\node[circle, draw, ultra thick, color = fusia!30, fill=fusia!30, minimum size = 32, minimum height = 32] at (7.1,-1.6) {};
\draw[thick, color=black, arrows = {-Latex[width=5pt]}] (5.95,-1.6)--(6.45,-1.6);

\node[text width=2cm] at (8.2,-0.08) {\color{cayenne} \small \texttt{48-bit}};
\node[text width=2cm] at (8.2,-0.42) {\color{cayenne} \small \texttt{Accumulator}};

\draw[thick, color=black] (7.1,-1.6)--(7.1,-2.25);
\draw[thick, color=black] (7.1,-1.6)--(6.55,-1.25);
\draw[thick, color=black] (7.1,-1.6)--(7.65,-1.25);

\node[text width=0.3cm] at (7.1,-1.25) {\color{black} \large \texttt{+}};
\node[text width=0.3cm] at (7.40,-1.85) {\color{black} \large \texttt{-}};

\node[text width=0.3cm] at (6.8,-1.85) {\color{black} \large \texttt{$\parallel$}};

%Mux
\draw[fill=scarl!30, color=scarl!30] (5.6,-4.4) -- (5.6,-2.8) -- (6.0,-3.1) -- (6.0,-4.1) -- cycle;

\draw[thick, color=black] (1.07,-4.8)--(4.8,-4.8);
\draw[thick, color=black] (4.8,-4.8)--(4.8,-3.1);
\draw[thick, color=black, arrows = {-Latex[width=5pt]}] (4.8,-3.1)--(5.6,-3.1);

\draw[thick, color=black] (6.0,-3.6)--(6.5,-3.6);
\draw[thick, color=black] (6.5,-3.6)--(6.5,-2.45);
\draw[thick, color=black, arrows = {-Latex[width=5pt]}] (6.5,-2.45)--(6.8,-2.15);

%Pattern Detector
\draw[fill=scarl!30, color=scarl!30] (7.6,-4.1) -- (7.6,-2.5) -- (8.4,-2.8) -- (8.4,-3.8) -- cycle;
\draw[thick, color=black] (7.1,-2.25)--(7.1,-3.3);
\draw[thick, color=black, arrows = {-Latex[width=5pt]}] (7.1,-3.3)--(7.6,-3.3);
\node[text width=0.3cm] at (8.0,-3.3) {\color{black} \large \texttt{=}};

% Final FFs
\node[rectangle, draw, ultra thick, color = ash, fill=ash, minimum size = 10, minimum height = 60] at (8.9,-2.45) {};

\draw[thick, color=black, arrows = {-Latex[width=5pt]}] (7.75,-1.6)--(8.7,-1.6);
\draw[thick, color=black, arrows = {-Latex[width=5pt]}] (8.4,-3.3)--(8.7,-3.3);

\draw[thick, color=black, arrows = {-Latex[width=5pt]}] (9.1,-2.45)--(9.8,-2.45);

\draw[thick, color=black] (9.4,-2.45)--(9.4,-4.8);
\draw[thick, color=black] (5.1,-4.8)--(9.4,-4.8);
\draw[thick, color=black] (5.1,-4.8)--(5.1,-4.1);

\draw[thick, color=black, arrows = {-Latex[width=5pt]}] (5.1,-4.1)--(5.6,-4.1);

\node[draw, color=white] at (9.4,-2.1) {\color{cayenne} \large \texttt{P}};

\end{tikzpicture}
\caption{Xilinx DSP48E1 architecture.}
\label{Fig21}
\vspace{-2ex}
\end{figure} 

The accuracy loss caused by the introduced approximation technique and fine-tuning is evaluated using the Alexnet \cite{a1} and VGG-16 \cite{a2} networks, and the Tiny ImageNet dataset. Compared to quantized fixed-point implementations of the Alexnet and VGG-16, our approximation and fine-tuning techniques can be implemented without loss of accuracy in almost all cases, while it causes slight accuracy losses in a few cases. 

After the multiplication packing and the parameter approximation, some pre-calculated values, which are used for single DSP - multiple parameter multiplication (SDMM), are stored on the on-chip ROM. Result of this, the actual parameter values are not necessary to store on the off-chip memory. Instead, the index values of the ROM are stored on the off-chip memory. This optimization provides up to 33\% compression on the off-chip memory without any hardware cost. The compression rate can be increased by up to 96.23\% and 97.03\% using Huffman coding and weight pruning for the Alexnet and VGG-16 networks, respectively. Also, the compression caused by the parameter representation change compensates the on-chip memory overhead caused by the ROM. It keeps on-chip memory size the same or less compared to traditional implementations. 

The main contributions of this paper are:
\begin{itemize}
\item We introduce a novel multiplication packing technique for multiplying one input variable with multiple constants as well as variables using a single DSP block (SDMM). In particular, we significantly reduce the usage of DSP blocks for FPGA implementations of CNN inference by packing the weights of a given CNN model using our technique.
\item We propose a novel approximation technique that allows using hardware resources efficiently without sacrificing accuracy. We achieve this by limiting the bit length of manipulated parameters.
\item We show a compression method that, significantly, reduces the off-chip memory transfers of parameters by using a ROM that stores pre-calculated results of our multiplication packing technique on on-chip memory. Our compression method guarantees a reduction of 33\%, 25\%, and 16.7\% for 8, 6, and 4-bit parameters when coupled with our approximation technique.
\item Finally, we design an area efficient systolic array architecture implementation. It uses 66.6\%, 75\%, and 83.3\% fewer DSP blocks for 8, 6, and 4-bit input variables compared to the baseline FPGA implementation. 
\end{itemize}

The rest of this paper is organized as follows. Background information and related works are summarized in Section 2. In Section 3, our technique is explained. Then, we describe our processing element architecture in Section 4. Section 5 explains the prototype systolic array architecture. Then, Section 6 presents the implementation results. Finally, the conclusion is given in Section 7. 

\section{Background and Related Work}
\subsection{DSP Blocks}

Different FPGA vendors provide built-in DSP blocks in their FPGAs. Although there are some differences between DSP architectures used by different FPGA manufacturers, all of these DSP blocks include the optimized hardware blocks required to handle MAC operations efficiently. Xilinx uses DSP48E1 and DSP48E2 blocks on different FPGAs. Fig. \ref{Fig21} shows the Xilinx DSP48E1 architecture, which has a 25-bit pre-adder, 25$\times$18 multiplier, and 48-bit accumulator. Xilinx DSP48E2 has a 27-bit pre-adder, 27$\times$18-bit multiplier, and 48-bit accumulator. 

DSP48 blocks can be configured differently for different type of operations. Xilinx DSP blocks can execute MAC operation using the multiplier and the accumulator available in the DSP48E1 (DSP48E2) as in \eqref{eq01}. Since it has a wide multiplier and accumulator, executing a MAC operation on the DSP block for reduced bit length fixed-point parameters cause underutilization issue. Using the 25$\times$18-bit multiplier for 8$\times$8-bit multiplication is an example of an underutilization problem. To fully utilize the DSP blocks, we develop an efficient multiplication packing technique, which can execute multiple parameter multiplication on a single DSP block (SDMM).

\begin{equation}
P = (A \cdot B) + C \label{eq01} 
\end{equation}

\subsection{Accelerator Design : CNN Use Case}

Research conducted on improving the utilization and/or performance of DSP blocks in the literature generally employs CNNs as an use-case, because CNN models require millions of MAC executions. We also implemented CNN inference on the systolic array architecture to evaluate the performance of our technique.

Convolution computations are still the main reason for the computational complexity of the CNN models, even if the state-of-the-art CNN models accommodate different layers such as pooling and activation. CNN models consist of multiple convolution layers, each containing many parallel convolution blocks. For example, VGG-16 has 13 different convolution layers, each with 64 to 512 parallel convolution channels. Also, each CNN kernel produces the result by summing the multiplication result of the CNN weight and input feature (I) with partial sums. As a result, state-of-the-art CNN models need to execute millions of multiply-accumulate (MAC) operations for classification. Table \ref{tab00} shows the required number of MAC operations for some CNN models.

\begin{table}[ht]
  \centering
  \caption{Number of MAC Required for Convolutions}
  \begin{tabular}{*{5}{c}}
    \toprule
    & Alexnet & VGG-16 & GoogleNet & MobileNet \\
    \midrule
    \# of MAC  & \multirow{2}{*}{666} & \multirow{2}{*}{15300} & \multirow{2}{*}{1233} & \multirow{2}{*}{568} \\
    (Millions) & \\
    \bottomrule
  \end{tabular}
  \label{tab00}
  \vspace{-2ex}
\end{table} 

Most of the CNN models are still trained on a GPU using floating-point weight values. On the other hand, the fact that CNN inference can be realized with negligible loss of accuracy using fixed-point reduced bit length CNN weights has made FPGA and ASIC designs a strong competitor of the GPU for CNN inference implementations \cite{f1} -\cite{f3}. Since FPGA and ASIC implementations offer high parallelism with low-power consumption, they are used to meet the intensive computing requirement of CNN models.

A systolic array (SA) architecture usually consists of a combination of processing elements (PE) in two dimensions, each of which can execute one MAC operation \cite{f4}. It provides a low-power and high-performance solution for matrix-matrix multiplications. Since SA architecture allows data movement between neighboring PEs only, it reduces the data movement cost significantly. Also, SA architecture enables data resue with different dataflow techniques. A SA architecture is widely used for CNN accelerators implemented on FPGA and ASIC \cite{f5} -\cite{f8}. Google TPU \cite{f7} and Xilinx xDNN \cite{f8} are two well-known examples where SA is used for CNN inference. We demonstrated the efficiency of our optimizations on the SA hardware.

\subsection{Related Work}
Previous publications in the literature about efficient use of DSP blocks can be classified into 3 groups; (1) optimizations targeting higher processing throughput, (2) optimizations enabling multiple parameter multiplications on a single DSP block, (3) novel DSP designs optimized for multiplication/MAC operations, which can be used in the future FPGA architectures. Our multiplication packing technique is in the second group.

In \cite{b4}, the authors investigated on the efficient use of cascade interconnection for DSP blocks to implement convolution kernels. They reorganized matrix-vector multiplication operations to make them cascade interconnection friendly, and implemented on the Xilinx UltraScale+ FPGA. Their FPGA implementation improved CNN inference latency, but it does not include an optimization to reduce the usage of DSP blocks. 

A novel method was introduced in \cite{b8} to perform two MAC operations per DSP block with the concatenation of two 8-bit parameters. This method reduced the DSP block usage by 50\% with the overhead of 11 LUTs and 12 FFs per MAC operation. The technique in \cite{b8} can only be used for 8-bit parameters, while our technique can support different parameter bit lengths such as 4, 6, and 8. Our multiplication packing technique uses less number of DSP blocks than the technique presented in \cite{b8}. Also, our technique has additional benefits like the compression on off-chip memory by parameter representation change.

An optimized DSP block (PIR-DSP) architecture was developed in \cite{b6} to implement reduced bit length fixed-point multiplications efficiently. They used run-time decomposable multiplier architecture and added a register file, which can also be configured as FIFO, to DSP block architecture. The PIR-DSP architecture increased the performance of the 9$\times$9-bit MAC operation 6 times compared to using the Xilinx DSP48E2 block. It also reduced energy consumption by 31\%. Similar to \cite{b6}, \cite{b7} developed a new DSP architecture based on the Intel Arria 10 DSP architecture. The DSP architecture given in \cite{b7} supports two simultaneous 8-bit parameter multiplications or four simultaneous 4-bit parameter multiplications. This increased the Arria 10 DSP size by 12\%. On the other hand, it reduced the DSP block usage by 15\% and 30\% for 8-bit and 4-bit parameters, respectively. Instead of using DSP architecture implemented on the state-of-the-art FPGAs, novel DSP architectures, which can be used for future FPGAs, were presented in \cite{b6} -\cite{b7}. Since FPGAs using these architectures are not available and the architectures of these DSP blocks are different from the existing ones, it is not possible to compare our multiplication packing technique with these implementations.

Xilinx provides the Deep Learning Processing Unit (DPU) in \cite{b9}. The DPU is an optimized programmable framework to implement deep learning algorithms efficiently on Xilinx FPGAs. Two different optimizations are available for MAC operations on a DPU. A DSP dual data rate (DDR) technique is supported in the DPU. The DSP-DDR technique, which is used to increase the throughput of the DSP blocks, enables using a high-frequency clock signal for DSP blocks. Also, the DPU offers two different resource allocation options (low-DSP and high-DSP). In case of a low-DSP usage option, DSP blocks are allocated only for multiplication operations. In case of a high-DSP usage option, DSP blocks are allocated for both multiplication and accumulation operations. The detailed comparison of our implementation with the DPU is given in Section 6. 

A method for constant multiplication packing was developed in \cite{c1}. This method reduced the bit-length of constants by a simple mathematical manipulation. Through the manipulation presented in \cite{c1}, the multiple constant multiplications can be executed on a single DSP block. This method carries out the manipulation of the constants using software and loads them to DSP blocks before runtime. As a result of this, only one tuple of constants can be executed on a DSP block during runtime. Although the technique developed in \cite{c1} works quite efficiently on hardware implementations used a small number of constants, as the number of constants to be multiplied is limited by the number of DSP blocks available in the FPGA, the increase in the number of constants significantly reduces the utilization efficiency of DSP blocks. Additionally, state-of-the-art DSP algorithms need much more multiplications than the number of DSP blocks available in the FPGA. As a consequence, it is not possible to implement state-of-the-art DSP algorithms in FPGA using the technique presented in \cite{c1}.

\section{Our Multiplication Packing Technique}

DSP blocks in FPGAs have wide multiplier and accumulator hardware considering bit-lengths used in various algorithms such as state-of-the-art CNN inference hardware implementations. The aforementioned causes the underutilization problem for DSP blocks, which can be solved by performing SDMM. Xilinx introduced a white-paper \cite{c2}, which offers a solution to implement two 8$\times$8 bit multiplications per DSP block using simple concatenation and 

\begin{algorithmic}[t]
\footnotesize
\DontPrintSemicolon
\vskip 0.9pt
\hrule
\vskip 5pt
 
\STATE \hskip-1.5em \textbf{Algorithm 1} Parameter manipulation
\vskip 5pt
\hrule
\vskip 5pt
\STATE \hskip-1em \textbf{input:} W
\STATE \hskip-1em \textbf{output:} MW, n, s
\STATE
\STATE \hskip-1em \textbf{function} manipulation(W)
\STATE
  \IF{$W > 0$}
    \WHILE{$\mod(W,2) = 0$}
      \STATE $s\gets s+1$\\
      \STATE $W\gets W\div 2$\
    \ENDWHILE
  \ENDIF
  \STATE
  $W\gets W-1$ \\
  \STATE
  \IF{$W > 0$}
    \WHILE{$\mod(W,2) = 0$}
      \STATE $n\gets n+1$\\
      \STATE $W\gets W\div 2$\
    \ENDWHILE
  \ENDIF
  \STATE
  \STATE $MW \gets W$
\STATE
\STATE \hskip-1em \textbf{end}
\vskip 5pt
\hrule
\vskip 8pt
%\EndFunction
\end{algorithmic} shift operations. The method given in \cite{c2} is only applicable for 2 8-bit parameters. Recently, Xilinx introduced an other white-paper \cite{c3}, which offers a solution to implement three 4$\times$4 bit multiplications per DSP block for DSP48E2.

Further, a mathematical manipulation method developed in \cite{c1} that can multiply multiple constants with one variable on a single DSP block. The method given in \cite{c1} is applicable only in cases where there are a small number of constants since the number of constants that can be multiplied by this manipulation is limited with the number of DSP blocks in the FPGA. Thus, it is not possible to implement state-of-the-art DSP algorithms, where millions of multiplications are required, using the method given in \cite{c1}.

We present an efficient multiplication packing technique, which allows the SDMM execution, based upon the constant manipulation technique in \cite{c1}. Firstly, in Section 3.1, we briefly explain the constant manipulation. Then, we introduce our novel parameter approximation technique in Section 3.2. Finally, our multiplication packing technique is described in Section 3.3. 

$I, W, MW,$ and $MW_A$ denote the input variable, fixed-point parameter, manipulated parameter, and manipulated approximate parameter. $v$ and $c$ denote the bit lengths of the input variable and the fixed-point parameter.

\subsection{Parameter Manipulation}
Xilinx DSP blocks have one multiplier and one accumulator hardware. As shown in \eqref{eq01}, DSP block can be configured for a MAC operation using one multiplier and one accumulator. We employ this option to configure our SDMM operation instead of a MAC operation. Accordingly, as shown in \eqref{eq1}, resources of the DSP block can be fully utilized by restructuring the multiplicand such that it consists of a multiplication and addition operations. This reduces the bit length of the multiplicand and shares the multiplication between the multiplier and accumulator hardware of the DSP block. The $n$ and $s$ values are determined to minimize MW as shown in Algorithm 1. 

\begin{equation}
W = {2^s} \cdot (1 + 2^n \cdot MW) \label{eq1} 
\end{equation}

The multiplication of the parameter (W) with the input variable (I) can be written as in \eqref{eq12}. Eq. \eqref{eq12} uses both the multiplier and accumulator hardware of the DSP block to perform a single multiplication operation. Eq. \eqref{eq12} makes multiple parameter multiplication on a single DSP block possible, since the multiplication operation is shared between the multiplier and accumulator hardware within the DSP block. The multiple multiplications are achieved by the concatenation of manipulated parameters for the multiplier and accumulator inputs of the DSP block.

\begin{equation}
P = W\cdot I = I \cdot {2^s} \cdot (1 + 2^n \cdot MW) \label{eq12} 
\end{equation}

\subsection{Our Novel Parameter Approximation Technique}

The number of parameters ($k$), which can be multiplied on a single DSP block depends on the bit-lengths of the input variable ($v$) and manipulated parameters ($c_i - (s_i + n_i)$). We set the $k$ number for the SDMM to 3, 4, and 6 for 8, 6, and 4-bit input variables, respectively.

The hardware implementation cost of the constant manipulation given in Algorithm 1 is higher than the implementing MAC operation using LUTs. As a solution, we design a Table Look Up based implementation, which stores the $n$ and $s$ values and the multiplicand input of the DSP block required for manipulated multiple parameter multiplication. The aforementioned Look-Up Table architecture is implemented on on-chip ROM and used as a dictionary for the SDMM. Depending on the input variable bit length, this Look-Up Table may need up to ($(2^v)^k$) different entries. This leads to MBs of on-chip ROM overhead, which is quite large for the on-chip storage capacity of the state-of-the-art FPGAs. As a solution, we introduce a novel parameter approximation technique to reduce the size of the on-chip ROM.

Our approximation technique gives a constraint over the mathematical manipulation equation given in \eqref{eq1}. The approximated version of \eqref{eq1} is shown in \eqref{eq9}. This approximation limits the bit length of $MW$ to a maximum of 3, regardless of the $W$. Although the value of $MW$ is limited to a maximum of 3 bits, 128 of 256 8-bit signed parameters can be implemented without any error thanks to $n$ and $s$. The other 128 values can be implemented with minor changes. The parameter approximation reduces the number of maximum different entries for the Look-Up Table to 8192, 16384, and 16384 for 8, 6, and 4-bit parameters, respectively. These numbers are found by software simulations. 

\begin{equation}
\setlength\abovedisplayskip{0pt}
W = {2^s} \cdot (1 + 2^n \cdot MW_A) \;\;\forall \: MW_A \in (0, 1, 3, 5, 7)  \label{eq9} 
\end{equation}

\subsection{Our Multiplication Packing Technique}

The multiplication packing technique separates MAC operation as multiplication and accumulation. 
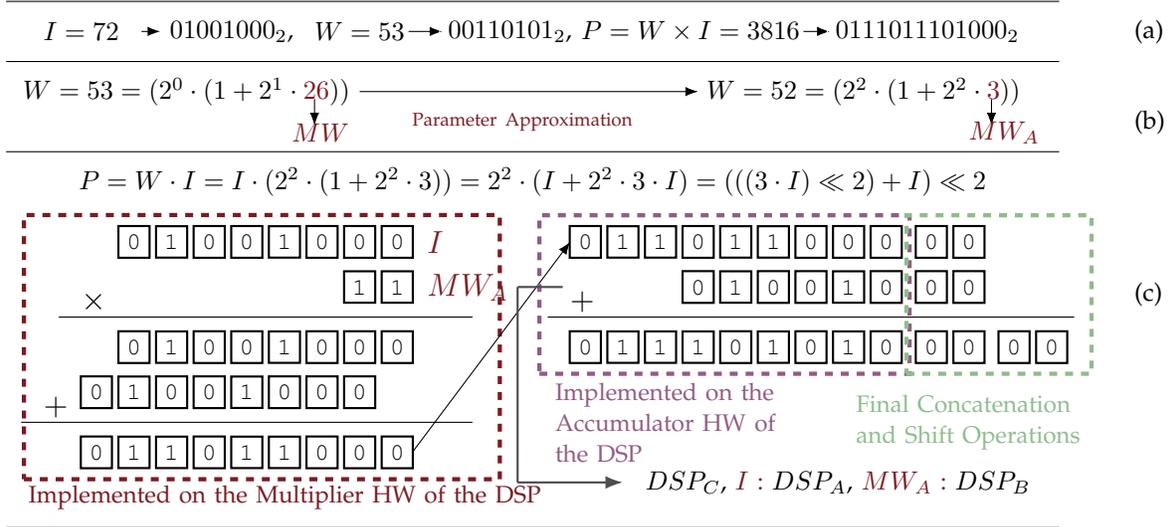
\begin{figure*}[ht]
\centering
\begin{tikzpicture}[PENode/.style={circle, draw, rectangle, thick, minimum size=10}]
\draw (-0.2,1.6) -- (13.8,1.6);

\node[draw, color=white] at (0.8,1.2) {\color{black} $I = 72$};
\draw[arrows = {-Latex}] (1.6,1.2)--(1.85,1.2);
\node[draw, color=white] at (2.8,1.2) {\color{black} $01001000_2$,};

\node[draw, color=white] at (4.5,1.2) {\color{black} $W = 53$};
\draw[arrows = {-Latex}] (5.15,1.2)--(5.6,1.2);
\node[draw, color=white] at (6.5,1.2) {\color{black} $00110101_2$,};

\node[draw, color=white] at (8.9,1.2) {\color{black} $P = W \times I = 3816$};
\draw[arrows = {-Latex}] (10.4,1.2)--(10.75,1.2);
\node[draw, color=white] at (12.05,1.2) {\color{black} $0111011101000_2$};

\node[text width = 0.4cm] at (15,1.2) {(a)};

\draw (-0.2,0.8) -- (13.8,0.8);

%\node[draw, color=white] at (0.8,0.4) {\color{black} $W = 53$};
\node[draw, color=white] at (2.2,0.4) {\color{black} $W = 53 = (2^0 \cdot (1+ 2^1 \cdot {\color{cayenne} 26}))$} ;
\draw[arrows = {-Latex}] (4.5,0.4)--(9,0.4);
\node[draw, color=white] at (11.2,0.4) {\color{black} $W = 52 = (2^2 \cdot (1+ 2^2 \cdot {\color{cayenne} 3}))$} ;
\draw[arrows = {-Latex}] (3.9,0.3)--(3.9,-0.05);
\node[text width = 0.5cm] at (3.85,-0.15) {$\color{cayenne} MW$};
\draw[arrows = {-Latex}] (12.9,0.3)--(12.9,-0.05);
\node[text width = 0.5cm] at (12.85,-0.15) {$\color{cayenne} MW_A$};

\node[text width = 4cm] at (7.2,0) {\color{cayenne} \scriptsize Parameter Approximation};

\node[text width = 0.4cm] at (15,0) {(b)};

\draw (-0.2,-0.4) -- (13.8,-0.4);

\node[draw, color=white] at (6.8,-0.8) {\color{black} $P = W \cdot I = I \cdot ( 2^2 \cdot (1+ 2^2 \cdot 3)) = 2^2 \cdot (I + 2^2 \cdot 3 \cdot I) = (((3 \cdot I) \ll 2) + I) \ll 2$} ;

\node[PENode] (00) at (1.5, -1.6) {\color{black}\texttt{\small{0}}};
\node[PENode] (01) at (2, -1.6) {\color{black}\texttt{\small{1}}};
\node[PENode] (02) at (2.5, -1.6) {\color{black}\texttt{\small{0}}};
\node[PENode] (03) at (3, -1.6) {\color{black}\texttt{\small{0}}};
\node[PENode] (04) at (3.5, -1.6) {\color{black}\texttt{\small{1}}};
\node[PENode] (05) at (4, -1.6) {\color{black}\texttt{\small{0}}};
\node[PENode] (06) at (4.5, -1.6) {\color{black}\texttt{\small{0}}};
\node[PENode] (07) at (5, -1.6) {\color{black}\texttt{\small{0}}};
\node[text width=0.4cm] at (5.6, -1.6) {\color{cayenne} \large $I$};

\node[PENode] (00) at (4.5, -2.2) {\color{black}\texttt{\small{1}}};
\node[PENode] (01) at (5, -2.2) {\color{black}\texttt{\small{1}}};
\node[text width=0.4cm] at (5.6, -2.2) {\color{cayenne} \large $MW_A$};

\draw (0.5,-2.6) -- (6,-2.6);

\node[text width=0.4cm] at (1, -2.4) {\color{black} \large $\times$};

\node[PENode] (00) at (1.5, -3.0) {\color{black}\texttt{\small{0}}};
\node[PENode] (01) at (2, -3.0) {\color{black}\texttt{\small{1}}};
\node[PENode] (00) at (2.5, -3.0) {\color{black}\texttt{\small{0}}};
\node[PENode] (01) at (3, -3.0) {\color{black}\texttt{\small{0}}};
\node[PENode] (00) at (3.5, -3.0) {\color{black}\texttt{\small{1}}};
\node[PENode] (01) at (4, -3.0) {\color{black}\texttt{\small{0}}};
\node[PENode] (00) at (4.5, -3.0) {\color{black}\texttt{\small{0}}};
\node[PENode] (01) at (5, -3.0) {\color{black}\texttt{\small{0}}};

\node[PENode] (00) at (1, -3.6) {\color{black}\texttt{\small{0}}};
\node[PENode] (01) at (1.5, -3.6) {\color{black}\texttt{\small{1}}};
\node[PENode] (00) at (2, -3.6) {\color{black}\texttt{\small{0}}};
\node[PENode] (01) at (2.5, -3.6) {\color{black}\texttt{\small{0}}};
\node[PENode] (00) at (3, -3.6) {\color{black}\texttt{\small{1}}};
\node[PENode] (01) at (3.5, -3.6) {\color{black}\texttt{\small{0}}};
\node[PENode] (00) at (4, -3.6) {\color{black}\texttt{\small{0}}};
\node[PENode] (01) at (4.5, -3.6) {\color{black}\texttt{\small{0}}};

\draw (0,-4.0) -- (6,-4.0);
\node[text width=0.4cm] at (0.5, -3.8) {\color{black} \large $+$};

\node[PENode] (00) at (1, -4.4) {\color{black}\texttt{\small{0}}};
\node[PENode] (00) at (1.5, -4.4) {\color{black}\texttt{\small{1}}};
\node[PENode] (01) at (2, -4.4) {\color{black}\texttt{\small{1}}};
\node[PENode] (00) at (2.5, -4.4) {\color{black}\texttt{\small{0}}};
\node[PENode] (01) at (3, -4.4) {\color{black}\texttt{\small{1}}};
\node[PENode] (00) at (3.5, -4.4) {\color{black}\texttt{\small{1}}};
\node[PENode] (01) at (4, -4.4) {\color{black}\texttt{\small{0}}};
\node[PENode] (00) at (4.5, -4.4) {\color{black}\texttt{\small{0}}};
\node[PENode] (01) at (5, -4.4) {\color{black}\texttt{\small{0}}};

\node[rectangle, draw, dashed, ultra thick, color = cayenne, minimum size = 180, minimum height = 100] at (3.2,-3) {};
\node[text width=7cm] at (3.6, -5) {\color{cayenne} \small Implemented on the Multiplier HW of the DSP};

\draw[arrows = {-Latex}, color=black] (5.2,-4.4)--(7.3,-1.6);

\node[PENode] (00) at (7.5, -1.6) {\color{black}\texttt{\small{0}}};
\node[PENode] (00) at (8, -1.6) {\color{black}\texttt{\small{1}}};
\node[PENode] (01) at (8.5, -1.6) {\color{black}\texttt{\small{1}}};
\node[PENode] (00) at (9, -1.6) {\color{black}\texttt{\small{0}}};
\node[PENode] (01) at (9.5, -1.6) {\color{black}\texttt{\small{1}}};
\node[PENode] (00) at (10, -1.6) {\color{black}\texttt{\small{1}}};
\node[PENode] (01) at (10.5, -1.6) {\color{black}\texttt{\small{0}}};
\node[PENode] (00) at (11, -1.6) {\color{black}\texttt{\small{0}}};
\node[PENode] (01) at (11.5, -1.6) {\color{black}\texttt{\small{0}}};

\node[PENode] (01) at (12.1, -1.6) {\color{black}\texttt{\small{0}}};
\node[PENode] (01) at (12.6, -1.6) {\color{black}\texttt{\small{0}}};

%\node[text width=3.6cm] at (14.8, -1.6) {\color{cayenne} \small $(MW \times I) \ll n$};

\node[PENode] (00) at (9, -2.2) {\color{black}\texttt{\small{0}}};
\node[PENode] (01) at (9.5, -2.2) {\color{black}\texttt{\small{1}}};
\node[PENode] (02) at (10, -2.2) {\color{black}\texttt{\small{0}}};
\node[PENode] (03) at (10.5, -2.2) {\color{black}\texttt{\small{0}}};
\node[PENode] (04) at (11, -2.2) {\color{black}\texttt{\small{1}}};
\node[PENode] (05) at (11.5, -2.2) {\color{black}\texttt{\small{0}}};
\node[PENode] (06) at (12.1, -2.2) {\color{black}\texttt{\small{0}}};
\node[PENode] (07) at (12.6, -2.2) {\color{black}\texttt{\small{0}}};
%\node[text width=0.4cm] at (14, -2.2) {\color{cayenne} \small $I$};

\draw (7,-2.6) -- (14,-2.6);

\node[text width=0.4cm] at (7.5, -2.4) {\color{black} \large $+$};

\node[PENode] (00) at (7.5, -3) {\color{black}\texttt{\small{0}}};
\node[PENode] (00) at (8, -3) {\color{black}\texttt{\small{1}}};
\node[PENode] (01) at (8.5, -3) {\color{black}\texttt{\small{1}}};
\node[PENode] (00) at (9, -3) {\color{black}\texttt{\small{1}}};
\node[PENode] (01) at (9.5, -3) {\color{black}\texttt{\small{0}}};
\node[PENode] (00) at (10, -3) {\color{black}\texttt{\small{1}}};
\node[PENode] (01) at (10.5, -3) {\color{black}\texttt{\small{0}}};
\node[PENode] (00) at (11, -3) {\color{black}\texttt{\small{1}}};
\node[PENode] (01) at (11.5, -3) {\color{black}\texttt{\small{0}}};

\node[PENode] (01) at (12.1, -3) {\color{black}\texttt{\small{0}}};
\node[PENode] (01) at (12.6, -3) {\color{black}\texttt{\small{0}}};

\node[PENode] (01) at (13.2, -3) {\color{black}\texttt{\small{0}}};
\node[PENode] (01) at (13.7, -3) {\color{black}\texttt{\small{0}}};

\node[rectangle, draw, dashed, ultra thick, color = fusia, minimum size = 140, minimum height = 60] at (9.35,-2.3) {};
\node[text width=3.4cm] at (8.8, -4.0) {\color{fusia} \small Implemented on the Accumulator HW of the DSP};

\node[rectangle, draw, dashed, ultra thick, color = scarl, minimum size = 70, minimum height = 60] at (13,-2.3) {};
\node[text width=3.4cm] at (12.8, -4.0) {\color{scarl} \small Final Concatenation and Shift Operations};

\draw[color = black!70, thick] (6.6,-2.2) -- (7.2,-2.2);
\draw[color = black!70, thick] (6.6,-2.2) -- (6.6,-4.8);
\draw[arrows = {-Latex}, color = black!70, thick] (6.6,-4.8)--(8.0,-4.8);

\node[text width=7cm] at (11.8, -4.8){\color{black} $DSP_C$, \color{cayenne} $I$ : \color{black} $DSP_A$, \color{cayenne} $MW_A$ : \color{black} $DSP_B$};

\draw (-0.2,-5.4) -- (13.8,-5.4);
\node[text width = 0.4cm] at (15,-2.3) {(c)};

\end{tikzpicture}
\caption{Numeric example of parameter manipulation, (a) Exact multiplication, (b) Parameter approximation, and (c) Manipulated multiplication.}
\label{fig10}
\end{figure*} Since the hardware cost of an accumulator is less than a multiplier, accumulation operation can be implemented using LUTs with a small hardware cost. Thanks to this, multiplier and accumulator hardware available in the DSP block can be used for the SDMM. The proposed multiplication technique is explained in 4 steps for simplicity; (1) multiplication packing for single parameter, (2) multiplication packing for signed input variable, (3) multiplication packing for multiple parameters, and (4) fine-tuning on parameter tuples.

\subsubsection{Multiplication Packing for Single Parameter}
The input variable (I) is multiplied with the approximate manipulated parameter as shown in \eqref{eq2}. Since the bit length of $MW_A$ is up to 3, multiple multiplications of the $MW_A$ with $I$ is possible by concatenating these multiplications on the multiplier of the DSP block. The remaining part of the multiplication with the parameter is executed on the accumulator hardware of the DSP block. As in \eqref{eq2}, the multiplication result ($MW_A \cdot I$) is shifted n-bits before the addition. This means that the least significant n-bits of the multiplication result are always zero. Hence, instead of performing addition on the least significant n-bits, the least significant n-bits of the I is concatenated directly to the addition result, which reduces the cost of the addition.

%\vspace{-2ex}
\begin{eqnarray}
\label{eq2}
W \cdot I & = & ({2^s} \cdot (1 + 2^n \cdot MW_A)) \cdot I \\ 
& = & (I + ((MW_A \cdot I) \ll n))\ll s \nonumber 
\end{eqnarray}

A numeric example is shown in Fig.~\ref{fig10}. Fig.~\ref{fig10} (a) shows the traditional multiplication of $I$ and $W$. As shown in Fig.~\ref{fig10} (b), the bit length of the $MW$ is reduced 5 to 2 with a small change in $W$ thanks to our approximation technique. The multiplication of the manipulated approximate parameter with the input variable is shown in Fig.~\ref{fig10} (c). The $MW_A$ is multiplied with $I$ using the multiplier hardware available in the DSP block. The MSBs of the multiplication result is accumulated with the MSBs of the $I$ using the accumulator hardware available in the DSP block. Finally, concatenation and shift operations are implemented at the output of the DSP block. 

\subsubsection{Signed Multiplication with Manipulation}

The aforementioned technique can perform a signed multiplication with the signed input variable. Still, some modifications are necessary on the multiplication packing technique for signed multiplication. Since multiple parameter multiplication is performed on a DSP block by concatenating multiple manipulated multiplications, the multiplier hardware in the DSP block performs multiplication by ignoring the addition of the sign extension part. Subsequently, the addition of sign extension bits are executed separately, and the result of the sign extension is added to the multiplication result in an intelligent way using accumulator hardware on the DSP block. That prevents the multiplication operation from using extra bits caused by the sign extension and makes possible multiplications of more parameters on a single DSP block.

In case of multiplication packing of exact parameters, sign extension bits for different parameters are calculated by using \eqref{eq6}. After that, $SEx$ should be concatenated with $I[v-1:n]$ to determine the value that that is used for the accumulation. 

%\vspace{-0.5ex}
\begin{equation}
SEx = (I[v-1] \cdot (2^{(m-s)} - W \cdot 2^{-s}))[(c-s-1):0]\label{eq6}
\end{equation}

Whereas, the proposed parameter approximation technique, significantly, simplifies the calculation of sign extension bits as shown \eqref{eq7}. Also, the value calculated in \eqref{eq7} is used for accumulation directly. It does not need extra concatenation. The $mask_{MW_A}$ equals to $111_2, 110_2, 100_2, 010_2,$ and $000_2$ for the $MW_A = 0, 1, 3, 5,$ and $7$, respectively. 

\begin{equation}
SEx_A = \{\left( mask_{MW_A} \; \& \; I[v-1] \right),(I \ggg n)\}\label{eq7}
\end{equation}

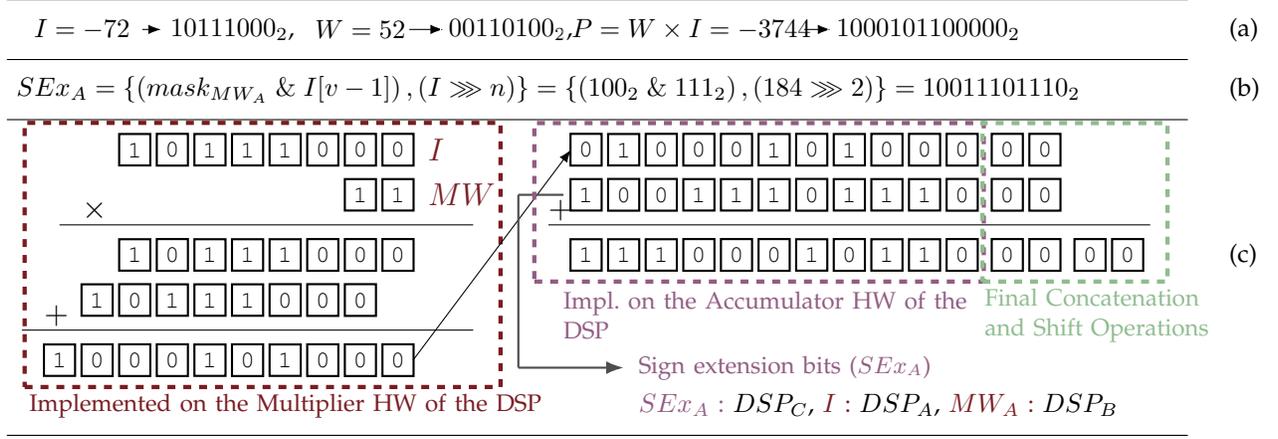
\begin{figure*}[ht]
\centering
\begin{tikzpicture}[PENode/.style={circle, draw, rectangle, thick, minimum size=10}]
\draw (-0.2,1.6) -- (15.5,1.6);

\node[draw, color=white] at (0.8,1.2) {\color{black} $I = -72$};
\draw[arrows = {-Latex}] (1.6,1.2)--(1.85,1.2);
\node[draw, color=white] at (2.8,1.2) {\color{black} $10111000_2$,};

\node[draw, color=white] at (4.5,1.2) {\color{black} $W = 52$};
\draw[arrows = {-Latex}] (5.15,1.2)--(5.6,1.2);
\node[draw, color=white] at (6.5,1.2) {\color{black} $00110100_2$,};

\node[draw, color=white] at (8.9,1.2) {\color{black} $P = W \times I = -3744$};
\draw[arrows = {-Latex}] (10.4,1.2)--(10.75,1.2);
\node[draw, color=white] at (12.05,1.2) {\color{black} $1000101100000_2$};

\node[text width = 0.4cm] at (16.25,1.2) {(a)};

\draw (-0.2,0.8) -- (15.5,0.8);

\node[draw, color=white] at (7,0.4) {\color{black} $SEx_A = \{\left( mask_{MW_A} \; \& \; I[v-1] \right),(I \ggg n)\} = \{\left( 100_2 \; \& \; 111_2 \right),(184 \ggg 2)\} = 10011101110_2$} ;

\node[text width = 0.4cm] at (16.25,0.4) {(b)};

\draw (-0.2,0.0) -- (15.5,0.0);

\node[PENode] (00) at (1.5, -0.4) {\color{black}\texttt{\small{1}}};
\node[PENode] (01) at (2, -0.4) {\color{black}\texttt{\small{0}}};
\node[PENode] (02) at (2.5, -0.4) {\color{black}\texttt{\small{1}}};
\node[PENode] (03) at (3, -0.4) {\color{black}\texttt{\small{1}}};
\node[PENode] (04) at (3.5, -0.4) {\color{black}\texttt{\small{1}}};
\node[PENode] (05) at (4, -0.4) {\color{black}\texttt{\small{0}}};
\node[PENode] (06) at (4.5, -0.4) {\color{black}\texttt{\small{0}}};
\node[PENode] (07) at (5, -0.4) {\color{black}\texttt{\small{0}}};
\node[text width=0.4cm] at (5.6, -0.4) {\color{cayenne} \large $I$};

\node[PENode] (00) at (4.5, -1.0) {\color{black}\texttt{\small{1}}};
\node[PENode] (01) at (5, -1.0) {\color{black}\texttt{\small{1}}};
\node[text width=0.4cm] at (5.6, -1.0) {\color{cayenne} \large $MW$};

\draw (0.5,-1.4) -- (6,-1.4);

\node[text width=0.4cm] at (1, -1.2) {\color{black} \large $\times$};

\node[PENode] (00) at (1.5, -1.8) {\color{black}\texttt{\small{1}}};
\node[PENode] (01) at (2, -1.8) {\color{black}\texttt{\small{0}}};
\node[PENode] (00) at (2.5, -1.8) {\color{black}\texttt{\small{1}}};
\node[PENode] (01) at (3, -1.8) {\color{black}\texttt{\small{1}}};
\node[PENode] (00) at (3.5, -1.8) {\color{black}\texttt{\small{1}}};
\node[PENode] (01) at (4, -1.8) {\color{black}\texttt{\small{0}}};
\node[PENode] (00) at (4.5, -1.8) {\color{black}\texttt{\small{0}}};
\node[PENode] (01) at (5, -1.8) {\color{black}\texttt{\small{0}}};

\node[PENode] (00) at (1, -2.4) {\color{black}\texttt{\small{1}}};
\node[PENode] (01) at (1.5, -2.4) {\color{black}\texttt{\small{0}}};
\node[PENode] (00) at (2, -2.4) {\color{black}\texttt{\small{1}}};
\node[PENode] (01) at (2.5, -2.4) {\color{black}\texttt{\small{1}}};
\node[PENode] (00) at (3, -2.4) {\color{black}\texttt{\small{1}}};
\node[PENode] (01) at (3.5, -2.4) {\color{black}\texttt{\small{0}}};
\node[PENode] (00) at (4, -2.4) {\color{black}\texttt{\small{0}}};
\node[PENode] (01) at (4.5, -2.4) {\color{black}\texttt{\small{0}}};

\draw (0,-2.8) -- (6,-2.8);
\node[text width=0.4cm] at (0.5, -2.6) {\color{black} \large $+$};

\node[PENode] (00) at (0.5, -3.2) {\color{black}\texttt{\small{1}}};
\node[PENode] (00) at (1, -3.2) {\color{black}\texttt{\small{0}}};
\node[PENode] (00) at (1.5, -3.2) {\color{black}\texttt{\small{0}}};
\node[PENode] (01) at (2, -3.2) {\color{black}\texttt{\small{0}}};
\node[PENode] (00) at (2.5, -3.2) {\color{black}\texttt{\small{1}}};
\node[PENode] (01) at (3, -3.2) {\color{black}\texttt{\small{0}}};
\node[PENode] (00) at (3.5, -3.2) {\color{black}\texttt{\small{1}}};
\node[PENode] (01) at (4, -3.2) {\color{black}\texttt{\small{0}}};
\node[PENode] (00) at (4.5, -3.2) {\color{black}\texttt{\small{0}}};
\node[PENode] (01) at (5, -3.2) {\color{black}\texttt{\small{0}}};

\node[rectangle, draw, dashed, ultra thick, color = cayenne, minimum size = 180, minimum height = 100] at (3.2,-1.8) {};
\node[text width=7cm] at (3.6, -3.8) {\color{cayenne} \small Implemented on the Multiplier HW of the DSP};

\draw[arrows = {-Latex}, color=black] (5.2,-3.2)--(7.3,-0.4);

\node[PENode] (00) at (7.5, -0.4) {\color{black}\texttt{\small{0}}};
\node[PENode] (00) at (8, -0.4) {\color{black}\texttt{\small{1}}};

\node[PENode] (00) at (8.5, -0.4) {\color{black}\texttt{\small{0}}};
\node[PENode] (00) at (9, -0.4) {\color{black}\texttt{\small{0}}};
\node[PENode] (01) at (9.5, -0.4) {\color{black}\texttt{\small{0}}};
\node[PENode] (00) at (10, -0.4) {\color{black}\texttt{\small{1}}};
\node[PENode] (01) at (10.5, -0.4) {\color{black}\texttt{\small{0}}};
\node[PENode] (00) at (11, -0.4) {\color{black}\texttt{\small{1}}};
\node[PENode] (01) at (11.5, -0.4) {\color{black}\texttt{\small{0}}};
\node[PENode] (00) at (12, -0.4) {\color{black}\texttt{\small{0}}};
\node[PENode] (01) at (12.5, -0.4) {\color{black}\texttt{\small{0}}};

\node[PENode] (01) at (13.1, -0.4) {\color{black}\texttt{\small{0}}};
\node[PENode] (01) at (13.6, -0.4) {\color{black}\texttt{\small{0}}};

\draw[color = black!70, thick] (6.6,-1.0) -- (7.2,-1.0);
\draw[color = black!70, thick] (6.6,-1.0) -- (6.6,-3.3);
\draw[arrows = {-Latex}, color = black!70, thick] (6.6,-3.3)--(8.0,-3.3);
\node[text width=5.6cm] at (11, -3.3) {\color{fusia} \small Sign extension bits ($SEx_A$)};

\node[PENode] (00) at (7.5, -1.0) {\color{black}\texttt{\small{1}}};
\node[PENode] (00) at (8, -1.0) {\color{black}\texttt{\small{0}}};
\node[PENode] (00) at (8.5, -1.0) {\color{black}\texttt{\small{0}}};
\node[PENode] (00) at (9, -1.0) {\color{black}\texttt{\small{1}}};
\node[PENode] (01) at (9.5, -1.0) {\color{black}\texttt{\small{1}}};
\node[PENode] (00) at (10, -1.0) {\color{black}\texttt{\small{1}}};
\node[PENode] (01) at (10.5, -1.0) {\color{black}\texttt{\small{0}}};
\node[PENode] (02) at (11, -1.0) {\color{black}\texttt{\small{1}}};
\node[PENode] (03) at (11.5, -1.0) {\color{black}\texttt{\small{1}}};
\node[PENode] (04) at (12, -1.0) {\color{black}\texttt{\small{1}}};
\node[PENode] (05) at (12.5, -1.0) {\color{black}\texttt{\small{0}}};
\node[PENode] (06) at (13.1, -1.0) {\color{black}\texttt{\small{0}}};
\node[PENode] (07) at (13.6, -1.0) {\color{black}\texttt{\small{0}}};
%\node[text width=0.4cm] at (14, -2.2) {\color{cayenne} \small $I$};

\draw (7,-1.4) -- (15,-1.4);

\node[text width=0.4cm] at (7.2, -1.2) {\color{black} \large $+$};

\node[PENode] (00) at (7.5, -1.8) {\color{black}\texttt{\small{1}}};
\node[PENode] (00) at (8, -1.8) {\color{black}\texttt{\small{1}}};
\node[PENode] (01) at (8.5, -1.8) {\color{black}\texttt{\small{1}}};
\node[PENode] (00) at (9, -1.8) {\color{black}\texttt{\small{0}}};
\node[PENode] (01) at (9.5, -1.8) {\color{black}\texttt{\small{0}}};
\node[PENode] (00) at (10, -1.8) {\color{black}\texttt{\small{0}}};
\node[PENode] (01) at (10.5, -1.8) {\color{black}\texttt{\small{1}}};
\node[PENode] (00) at (11, -1.8) {\color{black}\texttt{\small{0}}};
\node[PENode] (01) at (11.5, -1.8) {\color{black}\texttt{\small{1}}};

\node[PENode] (01) at (12.0, -1.8) {\color{black}\texttt{\small{1}}};
\node[PENode] (01) at (12.5, -1.8) {\color{black}\texttt{\small{0}}};

\node[PENode] (01) at (13.1, -1.8) {\color{black}\texttt{\small{0}}};
\node[PENode] (01) at (13.6, -1.8) {\color{black}\texttt{\small{0}}};

\node[PENode] (01) at (14.2, -1.8) {\color{black}\texttt{\small{0}}};
\node[PENode] (01) at (14.7, -1.8) {\color{black}\texttt{\small{0}}};

\node[rectangle, draw, dashed, ultra thick, color = fusia, minimum size = 170, minimum height = 60] at (9.8,-1.1) {};
\node[text width=5.6cm] at (10, -2.6) {\color{fusia} \small Impl. on the Accumulator HW of the DSP};

\node[rectangle, draw, dashed, ultra thick, color = scarl, minimum size = 70, minimum height = 60] at (14.0,-1.1) {};
\node[text width=3.4cm] at (14.5, -2.6) {\color{scarl} \small Final Concatenation and Shift Operations};

\node[text width=7cm] at (11.7, -3.8){\color{fusia} $SEx_A$ : \color{black} $DSP_C$, \color{cayenne} $I$ : \color{black} $DSP_A$, \color{cayenne} $MW_A$ : \color{black} $DSP_B$};

\draw (-0.2,-4.2) -- (15.5,-4.2);
\node[text width = 0.4cm] at (16.25,-1.8) {(c)};

\end{tikzpicture}
\caption{Numeric example of parameter manipulation (Signed input), (a) Signed multiplication, (b) Generation of sign extension bits, and (c) Signed multiplication using the multiplication packing.}
\label{fig11}
\end{figure*}

A numeric example of a signed multiplication with the proposed multiplication packing is given in Fig.~\ref{fig11}. Fig.~\ref{fig11} (a) shows the multiplication of the signed $I$ and $W$. Fig.~\ref{fig11} (b) generates the sign extension bits ($SEx_A$) using \eqref{eq7}. Fig.~\ref{fig11} (c) presents the signed multiplication.

\subsubsection{Manipulation for Multiple Parameters}

Firstly, all parameters to be multiplied on a single DSP block are manipulated using \eqref{eq9}. Afterward, depending on the input variable bit length, parameter tuples are generated for the SDMM. For example, each parameter tuple includes 3 parameters for the 8-bit input variable. The parameters within the given parameter tuple are concatenated as shown in \eqref{eq8}. The second row of the \eqref{eq8} is executed on the multiplier hardware available in the DSP block, while the third row is executed on the accumulator hardware available in the DSP block. Since the \eqref{eq8} includes sign extension calculations, it also supports signed multiplication. Final concatenation (with $I[n_i:0]$) and shift (with $s_i$) operations are not shown in \eqref{eq8} for simplicity. 

%\vspace{-2ex}
\begin{eqnarray}{C}
\label{eq8}
\scalebox{0.9}{$ \displaystyle \{W_k, ... W_2, W_1\} \cdot I = $} \\ 
\scalebox{0.9}{$ \displaystyle I \cdot \sum_{i=2}^{k} \left( MW_{A_i} \ll (i-1) \cdot (v + 3) \right) + MW_{A_1} \nonumber $} \\
\scalebox{0.9}{$ \displaystyle + \sum_{i=1}^{k} \left(SEx_{A_i} \ll (i-1) \cdot (v + 3) \right) \nonumber $} 
\end{eqnarray}

Eq. \eqref{eq8} is the simplified version of the exact version of the multiple manipulated parameter multiplication due to the novel parameter approximation technique. If the approximation technique is not used, the value of ($c_i - (n_i + s_i)$) would have to be calculated for each parameter within the parameter tuple, instead of 3 given in the second and third lines of \eqref{eq8}. The novel approximation ensures that the extra overhead caused by these calculations on the hardware is eliminated. In addition, it is necessary to use \eqref{eq6} instead of \eqref{eq7} for the sign extension, when the proposed approximation technique is not used. 

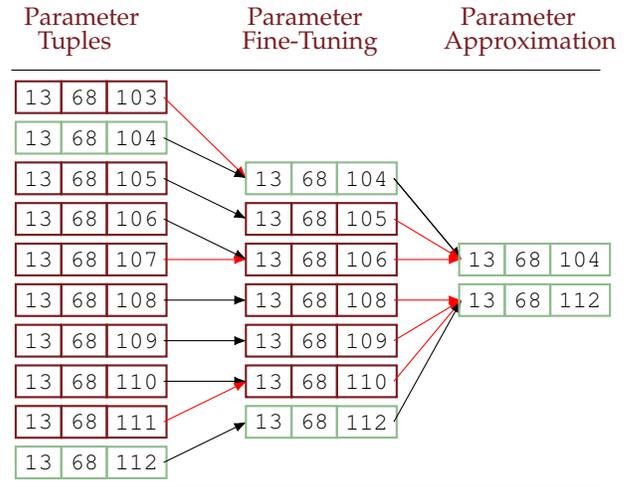
\begin{figure}%[ht]
\centering
\begin{tikzpicture}[PENode/.style={circle, draw, rectangle, thick, minimum size=10}, scale=0.9]
\draw (-2,0.4) -- (6.7,0.4);

\node[text width=2cm] at (-.7,1.2) {\color{cayenne} Parameter};
\node[text width=2cm] at (-0.5,0.8) {\color{cayenne} Tuples};

\node[text width=2cm] at (2.6,1.2) {\color{cayenne} Parameter};
\node[text width=2cm] at (2.52,0.8) {\color{cayenne} Fine-Tuning};

\node[text width=2cm] at (5.75,1.2) {\color{cayenne} Parameter};
\node[text width=2cm] at (5.5,0.8) {\color{cayenne} Approximation};

\node[PENode, color=cayenne] (00) at (-1.6, 0) {\color{black}\texttt{\small{13}}};
\node[PENode, color=cayenne] (01) at (-0.925, 0) {\color{black}\texttt{\small{68}}};
\node[PENode, color=cayenne] (02) at (-0.15, 0) {\color{black}\texttt{\small{103}}};

\node[PENode, color=scarl] (10) at (-1.6, -0.6) {\color{black}\texttt{\small{13}}};
\node[PENode, color=scarl] (11) at (-0.925, -0.6) {\color{black}\texttt{\small{68}}};
\node[PENode, color=scarl] (12) at (-0.15, -0.6) {\color{black}\texttt{\small{104}}};

\node[PENode, color=cayenne] (20) at (-1.6, -1.2) {\color{black}\texttt{\small{13}}};
\node[PENode, color=cayenne] (21) at (-0.925, -1.2) {\color{black}\texttt{\small{68}}};
\node[PENode, color=cayenne] (22) at (-0.15, -1.2) {\color{black}\texttt{\small{105}}};

\node[PENode, color=cayenne] (30) at (-1.6, -1.8) {\color{black}\texttt{\small{13}}};
\node[PENode, color=cayenne] (31) at (-0.925, -1.8) {\color{black}\texttt{\small{68}}};
\node[PENode, color=cayenne] (32) at (-0.15, -1.8) {\color{black}\texttt{\small{106}}};

\node[PENode, color=cayenne] (40) at (-1.6, -2.4) {\color{black}\texttt{\small{13}}};
\node[PENode, color=cayenne] (41) at (-0.925, -2.4) {\color{black}\texttt{\small{68}}};
\node[PENode, color=cayenne] (42) at (-0.15, -2.4) {\color{black}\texttt{\small{107}}};

\node[PENode, color=cayenne] (50) at (-1.6, -3) {\color{black}\texttt{\small{13}}};
\node[PENode, color=cayenne] (51) at (-0.925, -3) {\color{black}\texttt{\small{68}}};
\node[PENode, color=cayenne] (52) at (-0.15, -3) {\color{black}\texttt{\small{108}}};

\node[PENode, color=cayenne] (60) at (-1.6, -3.6) {\color{black}\texttt{\small{13}}};
\node[PENode, color=cayenne] (61) at (-0.925, -3.6) {\color{black}\texttt{\small{68}}};
\node[PENode, color=cayenne] (62) at (-0.15, -3.6) {\color{black}\texttt{\small{109}}};

\node[PENode, color=cayenne] (70) at (-1.6, -4.2) {\color{black}\texttt{\small{13}}};
\node[PENode, color=cayenne] (71) at (-0.925, -4.2) {\color{black}\texttt{\small{68}}};
\node[PENode, color=cayenne] (72) at (-0.15, -4.2) {\color{black}\texttt{\small{110}}};

\node[PENode, color=cayenne] (80) at (-1.6, -4.8) {\color{black}\texttt{\small{13}}};
\node[PENode, color=cayenne] (81) at (-0.925, -4.8) {\color{black}\texttt{\small{68}}};
\node[PENode, color=cayenne] (82) at (-0.15, -4.8) {\color{black}\texttt{\small{111}}};

\node[PENode, color=scarl] (90) at (-1.6, -5.4) {\color{black}\texttt{\small{13}}};
\node[PENode, color=scarl] (91) at (-0.925, -5.4) {\color{black}\texttt{\small{68}}};
\node[PENode, color=scarl] (92) at (-0.15, -5.4) {\color{black}\texttt{\small{112}}};

\draw[arrows = {-Latex}, color=red] (0.25,0)--(1.48,-1.2);
\draw[arrows = {-Latex}, color=black] (0.25,-0.6)--(1.48,-1.2);
\draw[arrows = {-Latex}, color=black] (0.25,-1.2)--(1.48,-1.8);
\draw[arrows = {-Latex}, color=black] (0.25,-1.8)--(1.48,-2.4);
\draw[arrows = {-Latex}, color=red] (0.25,-2.4)--(1.48,-2.4);
\draw[arrows = {-Latex}, color=black] (0.25,-3.0)--(1.48,-3.0);
\draw[arrows = {-Latex}, color=black] (0.25,-3.6)--(1.48,-3.6);
\draw[arrows = {-Latex}, color=black] (0.25,-4.2)--(1.48,-4.2);
\draw[arrows = {-Latex}, color=red] (0.25,-4.8)--(1.48,-4.2);
\draw[arrows = {-Latex}, color=black] (0.25,-5.4)--(1.48,-4.8);

\node[PENode, color=scarl] (23) at (1.8, -1.2) {\color{black}\texttt{\small{13}}};
\node[PENode, color=scarl] (24) at (2.475, -1.2) {\color{black}\texttt{\small{68}}};
\node[PENode, color=scarl] (25) at (3.25, -1.2) {\color{black}\texttt{\small{104}}};

\node[PENode, color=cayenne] (33) at (1.8, -1.8) {\color{black}\texttt{\small{13}}};
\node[PENode, color=cayenne] (34) at (2.475, -1.8) {\color{black}\texttt{\small{68}}};
\node[PENode, color=cayenne] (35) at (3.25, -1.8) {\color{black}\texttt{\small{105}}};

\node[PENode, color=cayenne] (43) at (1.8, -2.4) {\color{black}\texttt{\small{13}}};
\node[PENode, color=cayenne] (44) at (2.475, -2.4) {\color{black}\texttt{\small{68}}};
\node[PENode, color=cayenne] (45) at (3.25, -2.4) {\color{black}\texttt{\small{106}}};

\node[PENode, color=cayenne] (53) at (1.8, -3) {\color{black}\texttt{\small{13}}};
\node[PENode, color=cayenne] (54) at (2.475, -3) {\color{black}\texttt{\small{68}}};
\node[PENode, color=cayenne] (55) at (3.25, -3) {\color{black}\texttt{\small{108}}};

\node[PENode, color=cayenne] (63) at (1.8, -3.6) {\color{black}\texttt{\small{13}}};
\node[PENode, color=cayenne] (64) at (2.475, -3.6) {\color{black}\texttt{\small{68}}};
\node[PENode, color=cayenne] (65) at (3.25, -3.6) {\color{black}\texttt{\small{109}}};

\node[PENode, color=cayenne] (73) at (1.8, -4.2) {\color{black}\texttt{\small{13}}};
\node[PENode, color=cayenne] (74) at (2.475, -4.2) {\color{black}\texttt{\small{68}}};
\node[PENode, color=cayenne] (75) at (3.25, -4.2) {\color{black}\texttt{\small{110}}};

\node[PENode, color=scarl] (83) at (1.8, -4.8) {\color{black}\texttt{\small{13}}};
\node[PENode, color=scarl] (84) at (2.475, -4.8) {\color{black}\texttt{\small{68}}};
\node[PENode, color=scarl] (85) at (3.25, -4.8) {\color{black}\texttt{\small{112}}};

\draw[arrows = {-Latex}, color=black] (3.65,-1.2)--(4.65,-2.4);

\draw[arrows = {-Latex}, color=black] (3.65,-1.2)--(4.65,-2.4);
\draw[arrows = {-Latex}, color=red] (3.65,-1.8)--(4.65,-2.4);
\draw[arrows = {-Latex}, color=red] (3.65,-2.4)--(4.65,-2.4);

\draw[arrows = {-Latex}, color=red] (3.65,-3.0)--(4.65,-3.0);
\draw[arrows = {-Latex}, color=red] (3.65,-3.6)--(4.65,-3.0);
\draw[arrows = {-Latex}, color=red] (3.65,-4.2)--(4.65,-3.0);
\draw[arrows = {-Latex}, color=black] (3.65,-4.8)--(4.65,-3.0);

\node[PENode, color=scarl] (46) at (4.95, -2.4) {\color{black}\texttt{\small{13}}};
\node[PENode, color=scarl] (47) at (5.625, -2.4) {\color{black}\texttt{\small{68}}};
\node[PENode, color=scarl] (48) at (6.4, -2.4) {\color{black}\texttt{\small{104}}};

\node[PENode, color=scarl] (56) at (4.95, -3) {\color{black}\texttt{\small{13}}};
\node[PENode, color=scarl] (57) at (5.625, -3) {\color{black}\texttt{\small{68}}};
\node[PENode, color=scarl] (58) at (6.4, -3) {\color{black}\texttt{\small{112}}};

\draw (-2,-5.8) -- (6.7,-5.8);

\end{tikzpicture}
\caption{Numeric example of the parameter approximation and fine-tuning (Red Arrow : Approximated).}
\label{fig12}
%\vspace{-2ex}
\end{figure}

\subsubsection{Fine-tuning of Parameter Tuples}

The input variable bit length ($v$) is the major term to determine the number of parameters that can be multiplied on a single DSP block. Whereas, depending on the parameter values, it is not always possible to perform a SDMM with this number. Thus, a fine-tuning of parameters is performed to guarantee that the number of parameter multiplication per DSP block is fixed. Unlike the traditional techniques in which each parameter is independently tuned, our technique performs fine-tuning on parameter tuples. 

Our fine-tuning technique consists of three steps. First, the number of parameters that can be multiplied on a single DSP block is determined based on the bit-lengths of the input variable and the parameters. 3, 4, and 6 parameters can be multiplied on a single DSP block for the 8, 6, and 4-bit input variables (I), respectively. Later, all of the parameter tuples that can be multiplied on a single DSP block is determined using the fixed number found in the first step. Finally, the parameter tuple, which cannot be multiplied on a single DSP block, is replaced by the closest parameter tuple in the set determined in the second step. The Bray-Curtis distance formula, which is given in \eqref{eq99}, is used to determine the closest parameter tuple. 

\begin{eqnarray}{C}
\label{eq99}
u : 1D\;\;Array,\;\;\;\;v : 1D\;\;Array \\ 
BC = \sum \mid \mid u_i \mid  - \mid v_i \mid \mid / \sum \mid u_i + v_i \mid \nonumber 
\end{eqnarray}
%\vspace{0.3ex}

\begin{table*}%[htbp]
  \centering
  \renewcommand{\arraystretch}{1.2}
  \caption{Error Increase (\%) Caused by The Approximation Technique (CNN Use Case)}
  \begin{tabular}{*{10}{c}}
    \hline
    \multirow{2}{*}{CNN Model} & \multicolumn{9}{c}{(W,I) : Bit Lengths of CNN Weight and Input Variable} \\
    \cline{2-10}
    & (8,8) & (8,6) & (8,4) & (6,8) & (6,6) & (6,4) & (4,8) & (4,6) & (4,4) \\
    \hline
    Alexnet & -0.18 & -0.21 & -0.38 & 0.08 & 0.30 & 0.04 & 0.00 & 0.00 & 0.00 \\
  \hline
    VGG-16 & 0.01 & 0.05 & -0.19 & 0.04 & -0.31 & -0.06 & 0.00 & 0.00 & 0.00 \\
    \hline
  \end{tabular}
  \label{tab10}
%  \vspace{-1.6ex}
\end{table*}

\begin{figure*}%[h]
\centering
\begin{tikzpicture}[INode/.style={circle, draw, square, thick, color=salmon, fill=salmon, minimum size=8}, 
                    PENode/.style={circle, draw, square, thick, color=cayenne, minimum size=24}]

\node[text width=3cm] at (0.4,0.5) {\color{cayenne} \small \texttt{Parameter Decomp.}};
\node[text width=3cm] at (5,0.1) {\color{cayenne} \small \texttt{Multiple Parameter Multiplication}};
\node[text width=3cm] at (10.5,0.1) {\color{cayenne} \small \texttt{Post-Processing}};

\node[text width=3cm] at (15,0.45) {\color{cayenne} \small \texttt{Accumulation}};

\node[text width=0.5cm] at (-0.4,-0.4) {\color{cayenne} \scriptsize \texttt{I}};
\node[text width=0.5cm] at (-0.4,-0.9) {\color{cayenne} \scriptsize \texttt{M3}};
\node[text width=1.7cm] at (-0.4,-1.45) {\color{cayenne} \scriptsize \texttt{\{3\{I[v]\}\}}};

\node[text width=0.5cm] at (-0.4,-1.9) {\color{cayenne} \scriptsize \texttt{I}};
\node[text width=0.5cm] at (-0.4,-2.4) {\color{cayenne} \scriptsize \texttt{M2}};
\node[text width=1.7cm] at (-0.4,-2.95) {\color{cayenne} \scriptsize \texttt{\{3\{I[v]\}\}}};

\node[text width=0.5cm] at (-0.4,-3.4) {\color{cayenne} \scriptsize \texttt{I}};
\node[text width=0.5cm] at (-0.4,-3.9) {\color{cayenne} \scriptsize \texttt{M1}};
\node[text width=1.7cm] at (-0.4,-4.45) {\color{cayenne} \scriptsize \texttt{\{3\{I[v]\}\}}};

\node[rectangle, draw, ultra thick, color = ash, fill=ash, minimum size = 16, minimum height = 16] at (0.3,-0.4) {};
\draw[thick, color=black] (-0.3,-0.4)--(0,-0.4);
\node[text width = .37cm] at (0.3,-0.4) {\color{black} \scriptsize \textbf{$\ggg$}};
\draw[thick, color=black, arrows = {-Latex[width=5pt]}] (0.6,-0.4)--(1.1,-0.4);

\node[and gate US, draw, cayenne!50, fill=cayenne!50, line width = 0.3pt, minimum height = 0.3cm] (A) at (0.3,-1.15) {};
\draw[thick, color=black] (-0.3,-1.05)--(0.02,-1.05);
\draw[thick, color=black] (-0.3,-1.25)--(0.02,-1.25);
\draw[thick, color=black, arrows = {-Latex[width=5pt]}] (0.62,-1.15)--(1.1,-1.15);

\node[rectangle, draw, ultra thick, color = ash, fill=ash, minimum size = 16, minimum height = 16] at (0.3,-1.9) {};
\draw[thick, color=black] (-0.3,-1.9)--(0,-1.9);
\node[text width = .37cm] at (0.3,-1.9) {\color{black} \scriptsize \textbf{$\ggg$}};
\draw[thick, color=black, arrows = {-Latex[width=5pt]}] (0.6,-1.9)--(1.1,-1.9);

\node[and gate US, draw, cayenne!50, fill=cayenne!50, line width = 0.3pt, minimum height = 0.3cm] (A) at (0.3,-2.65) {};
\draw[thick, color=black] (-0.3,-2.55)--(0.02,-2.55);
\draw[thick, color=black] (-0.3,-2.75)--(0.02,-2.75);
\draw[thick, color=black, arrows = {-Latex[width=5pt]}] (0.62,-2.65)--(1.1,-2.65);

\node[rectangle, draw, ultra thick, color = ash, fill=ash, minimum size = 16, minimum height = 16] at (0.3,-3.4) {};
\draw[thick, color=black] (-0.3,-3.4)--(0,-3.4);
\node[text width = .37cm] at (0.3,-3.4) {\color{black} \scriptsize \textbf{$\ggg$}};
\draw[thick, color=black, arrows = {-Latex[width=5pt]}] (0.6,-3.4)--(1.1,-3.4);

\node[and gate US, draw, cayenne!50, fill=cayenne!50, line width = 0.3pt, minimum height = 0.3cm] (A) at (0.3,-4.15) {};
\draw[thick, color=black] (-0.3,-4.05)--(0.02,-4.05);
\draw[thick, color=black] (-0.3,-4.25)--(0.02,-4.25);
\draw[thick, color=black, arrows = {-Latex[width=5pt]}] (0.62,-4.15)--(1.1,-4.15);

\node[rectangle, draw, ultra thick, color = cayenne!30, fill=cayenne!30, minimum size = 13, minimum height = 130] at (1.3,-2.275) {};

\node[text width = 2cm, rotate=90] at (1.3,-2.275) {\scriptsize \texttt{Concatenation}};

\draw[thick, color=black] (1.55,-2.275)--(2,-2.275);
\draw[thick, color=black] (2,-2.275)--(2,-3.45);
\draw[thick, color=black] (2,-3.45)--(2.94,-3.45);

\node[text width=0.5cm] at (3.2,-1.3) {\color{cayenne} \large \texttt{B}};
\node[text width=0.5cm] at (3.2,-2.5) {\color{cayenne} \large \texttt{A}};
\node[text width=0.5cm] at (3.2,-3.7) {\color{cayenne} \large \texttt{C}};

% DSP INPUT FFs
\node[rectangle, draw, ultra thick, color = ash, fill=ash, minimum size = 8, minimum height = 16] at (3.5,-1.05) {};
\draw[thick, color=black] (2.43,-1.05)--(3.33,-1.05);

\node[rectangle, draw, ultra thick, color = ash, fill=ash, minimum size = 8, minimum height = 16] at (3.5,-2.25) {};
\draw[thick, color=black] (2.43,-2.25)--(3.33,-2.25);

\node[rectangle, draw, ultra thick, color = ash, fill=ash, minimum size = 8, minimum height = 16] at (3.5,-3.45) {};
\draw[thick, color=black] (2.43,-3.45)--(3.33,-3.45);

% DSP Multiplier
\node[circle, draw, ultra thick, color = fusia!30, fill=fusia!30, minimum size = 12, minimum height = 12] at (4.5,-1.65) {\color{black} \large \texttt{$\times$}};

\draw[thick, color=black, arrows = {-Latex[width=5pt]}] (3.95,-1.05)--(4.25,-1.35);
\draw[thick, color=black] (3.67,-1.05)--(3.95,-1.05);

\draw[thick, color=black, arrows = {-Latex[width=5pt]}] (3.95,-2.25)--(4.25,-1.95);
\draw[thick, color=black] (3.67,-2.25)--(3.95,-2.25);

% DSP Accumulator
\node[circle, draw, ultra thick, color = fusia!30, fill=fusia!30, minimum size = 5, minimum height = 5] at (5.8,-2.25) {\color{black} \large \texttt{$+$}};

\draw[thick, color=black, arrows = {-Latex[width=5pt]}] (5.25,-1.65)--(5.55,-1.95);
\draw[thick, color=black] (4.9,-1.65)--(5.25,-1.65);

\draw[thick, color=black, arrows = {-Latex[width=5pt]}] (5.25,-3.45)--(5.55,-2.55);
\draw[thick, color=black] (3.67,-3.45)--(5.25,-3.45);

\node[rectangle, draw, ultra thick, color = ash, fill=ash, minimum size = 8, minimum height = 16] at (6.85,-2.25) {};
\draw[thick, color=black, arrows = {-Latex[width=5pt]}] (6.2,-2.25)--(6.7,-2.25);
\draw[thick, color=black] (7,-2.25)--(8.5,-2.25);

\draw[thick, color=black] (8.5,-0.75)--(8.5,-3.75);

\draw[thick, color=black] (8.5,-3.75)--(9.5,-3.75);
\node[rectangle, draw, ultra thick, color = cayenne!30, fill=cayenne!30, minimum size = 16, minimum height = 16] at (9.8,-3.9) {\color{black} \scriptsize \texttt{C}};
\draw[thick, color=black] (8.8,-4.05)--(9.5,-4.05);
\node[text width=0.5cm] at (8.78,-4.2) {\tiny \color{cayenne} {\texttt{I[n1-1:0]}}};

\draw[thick, color=black] (8.5,-2.25)--(9.5,-2.25);
\node[rectangle, draw, ultra thick, color = cayenne!30, fill=cayenne!30, minimum size = 16, minimum height = 16] at (9.8,-2.4) {\color{black} \scriptsize \texttt{C}};
\draw[thick, color=black] (8.8,-2.55)--(9.5,-2.55);
\node[text width=0.5cm] at (8.78,-2.7) {\tiny \color{cayenne} \texttt{I[n2-1:0]}};

\draw[thick, color=black] (8.5,-0.75)--(9.5,-0.75);
\node[rectangle, draw, ultra thick, color = cayenne!30, fill=cayenne!30, minimum size = 16, minimum height = 16] at (9.8,-0.9) {\color{black} \scriptsize \texttt{C}};
\draw[thick, color=black] (8.8,-1.05)--(9.5,-1.05);
\node[text width=0.5cm] at (8.78,-1.2) {\tiny \color{cayenne} \texttt{I[n3-1:0]}};

\node[rectangle, draw, ultra thick, color = ash, fill=ash, minimum size = 16, minimum height = 16] at (11,-0.9) {\color{black} $\ll$};
\node[rectangle, draw, ultra thick, color = ash, fill=ash, minimum size = 16, minimum height = 16] at (11,-2.4) {\color{black} $\ll$};
\node[rectangle, draw, ultra thick, color = ash, fill=ash, minimum size = 16, minimum height = 16] at (11,-3.9) {\color{black} $\ll$};

\draw[thick, color=black, arrows = {-Latex[width=5pt]}] (10.1,-0.9)--(10.7,-0.9);
\draw[thick, color=black, arrows = {-Latex[width=5pt]}] (10.1,-2.4)--(10.7,-2.4);
\draw[thick, color=black, arrows = {-Latex[width=5pt]}] (10.1,-3.9)--(10.7,-3.9);

\node[rectangle, draw, ultra thick, color = scarl!30, fill=scarl!30, minimum size = 16, minimum height = 16] at (12.2,-0.9) {\color{black} \scriptsize \texttt{S}};
\node[rectangle, draw, ultra thick, color = scarl!30, fill=scarl!30, minimum size = 16, minimum height = 16] at (12.2,-2.4) {\color{black} \scriptsize \texttt{S}};
\node[rectangle, draw, ultra thick, color = scarl!30, fill=scarl!30, minimum size = 16, minimum height = 16] at (12.2,-3.9) {\color{black} \scriptsize \texttt{S}};

\draw[thick, color=black, arrows = {-Latex[width=5pt]}] (11.3,-0.9)--(11.9,-0.9);
\draw[thick, color=black, arrows = {-Latex[width=5pt]}] (11.3,-2.4)--(11.9,-2.4);
\draw[thick, color=black, arrows = {-Latex[width=5pt]}] (11.3,-3.9)--(11.9,-3.9);

\node[circle, draw, ultra thick, color = fusia!30, fill=fusia!30, minimum size = 5, minimum height = 5] at (14.5,-0.9) {\color{black} \large \texttt{$+$}};
\node[circle, draw, ultra thick, color = fusia!30, fill=fusia!30, minimum size = 5, minimum height = 5] at (14.5,-2.4) {\color{black} \large \texttt{$+$}};
\node[circle, draw, ultra thick, color = fusia!30, fill=fusia!30, minimum size = 5, minimum height = 5] at (14.5,-3.9) {\color{black} \large \texttt{$+$}};

\draw[thick, color=black, arrows = {-Latex[width=5pt]}] (12.5,-0.9)--(14.15,-0.9);
\draw[thick, color=black, arrows = {-Latex[width=5pt]}] (12.5,-2.4)--(14.15,-2.4);
\draw[thick, color=black, arrows = {-Latex[width=5pt]}] (12.5,-3.9)--(14.15,-3.9);

\draw[thick, color=black, arrows = {-Latex[width=5pt]}] (13.95,-0.25)--(14.25,-0.65);
\draw[thick, color=black] (13.1,-0.25)--(13.95,-0.25);
\draw[thick, color=black, arrows = {-Latex[width=5pt]}] (14.9,-0.9)--(16,-0.9);
\node[text width=0.5cm] at (13.4,0) {\scriptsize \color{cayenne} {\texttt{PSum3}}};

\draw[thick, color=black, arrows = {-Latex[width=5pt]}] (13.95,-1.85)--(14.25,-2.15);
\draw[thick, color=black] (13.1,-1.85)--(13.95,-1.85);
\draw[thick, color=black, arrows = {-Latex[width=5pt]}] (14.9,-2.4)--(16,-2.4);
\node[text width=0.5cm] at (13.4,-1.6) {\scriptsize \color{cayenne} {\texttt{PSum2}}};

\draw[thick, color=black, arrows = {-Latex[width=5pt]}] (13.95,-4.45)--(14.25,-4.15);
\draw[thick, color=black] (13.1,-4.45)--(13.95,-4.45);
\draw[thick, color=black, arrows = {-Latex[width=5pt]}] (14.9,-3.9)--(16,-3.9);
\node[text width=0.5cm] at (13.4,-4.7) {\scriptsize \color{cayenne} {\texttt{PSum1}}};

\node[text width=0.5cm] at (15.2,-3.6) {\scriptsize \color{cayenne} {\texttt{MACOut1}}};
\node[text width=0.5cm] at (15.2,-2.1) {\scriptsize \color{cayenne} {\texttt{MACOut2}}};
\node[text width=0.5cm] at (15.2,-0.6) {\scriptsize \color{cayenne} {\texttt{MACOut3}}};

\node[rectangle, draw, dashed, ultra thick, color = cayenne, minimum size = 90, minimum height = 140] at (0.25,-2.3) {};

\node[rectangle, draw, dashed, ultra thick, color = cayenne, minimum size = 140, minimum height = 100] at (5.2,-2.3) {};
\node[rectangle, draw, dashed, ultra thick, color = cayenne, minimum size = 130, minimum height = 120] at (10.5,-2.4) {};
\node[rectangle, draw, dashed, ultra thick, color = cayenne, minimum size = 90, minimum height = 147] at (14.6,-2.35) {};

\draw[thick, color=black] (7.35, -2.3) -- (7.45, -2.2) node[anchor=north] {};
\node[text width=0.5cm] at (7.5,-2.1) {\scriptsize \texttt{33}};

\end{tikzpicture}
\caption{Our PE architecture.}
\label{fig77}
%  \vspace{-2ex}
\end{figure*}
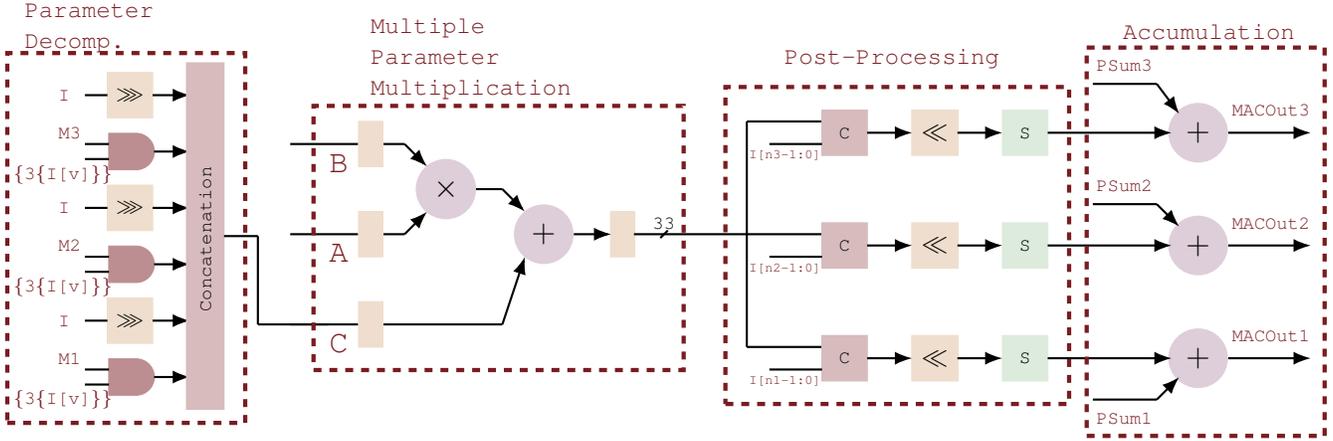 

A numeric example of the fine-tuning and parameter approximation is shown in Fig.~\ref{fig12}. Ten different parameter tuples are shown in the left column of the Fig.~\ref{fig12}. The parameter tuples, which cannot be multiplied on a single DSP block, are tuned using \eqref{eq99}. The fine-tuning step reduces the number of unique parameter tuples to seven. Furthermore, according to the approximation rule presented in \eqref{eq9}, some of the parameters are approximated. The parameter approximation reduced the number of unique parameter tuples to two. 

Accuracy of the parameter approximation and fine-tuning techniques for 4, 6, and 8-bit fixed-point parameters and 4, 6, and 8-bit input variables (I) were evaluated on the Alexnet and VGG-16 networks. Validation images of the Tiny ImageNet were used to measure accuracy results. The error increase caused by the approximation and fine-tuning is reported in Table \ref{tab10} compared to a quantized implementation of the Alexnet and VGG-16. As shown in Table \ref{tab10}, the parameter approximation and fine-tuning can decrease the error in some cases, while it causes a slight increase in the error in other cases. Since Eq. \eqref{eq9} can implement signed parameters smaller than 6-bits without any error, multiplication of parameters smaller than 6-bits can be performed without any error using the parameter approximation technique.

\section{Processing Element (PE) Architecture}

Our PE architecture is different from the traditional PE architectures implemented for MAC operation. Traditional PE implementations can execute one MAC operation per DSP block using the multiplier and accumulator hardware available in the DSP block. Our PE architecture is customized for the SDMM execution. The multiplication and the accumulation operations are executed separately in the implemented PE architecture. The multiplier and accumulator hardware available in the DSP block is configured to execute multiple parameter multiplications. Corresponding accumulations of these multiple parameter multiplications are executed using the LUTs available in the FPGA. This brings up two advantages. First, a wide multiplier hardware available in the DSP block can be fully-utilized for the reduced bit-length parameters. Second, it enables an efficient resource sharing in FPGAs with a limited number of DSP resources.

The PE architecture is shown in Fig.~\ref{fig77} for the 8-bit parameters and the 3 parameter multiplications per DSP block. The PE architecture is used by scaling for different bit lengths. This PE architecture consists of four parts; parameter decompression, multiple parameter multiplication, post-processing, and accumulation. 

In the parameter decompression part, the parameter decompression is performed using the ROM output. Since the calculation of the multiplicand input of the DSP block is independent of the input variable as shown in the second line of the \eqref{eq8}, the multiplicands ('A' input of the DSP block) are stored in the ROM for different parameter tuples. Also, $n$ and $s$ values for each parameter within the parameter tuple are stored in the ROM. The most significant 24 bits of the output of the ROM, which stores the multiplicand given in the second line of the \eqref{eq8}, are connected to the 'A' input of the DSP block directly. The least significant bits of the WROM output is used to store the shift values required to generate the 'C' input of the DSP block. The 'C' input of the DSP block is generated using the parameter decompression part shown in the Fig.~\ref{fig77}. Masks (M1, M2, and M3), which are used to generate the sign extension bits, are stored in the LUTs, and only 30 6-input LUTs are necessary for the entire hardware. The parameter decompression hardware needs 35 LUTs for each 3 parameter multiplications (for 8-bit fixed-point parameters). Also, the LUT overhead caused by the parameter decompression for the entire hardware is reported in Section VI.

Unlike the traditional PE implementations, SDMM (3 multiplications per PE for 8-bit fixed-point parameters) is executed in the second part of the design. This achieved using the multiplier and accumulator hardware available in the DSP block. 

The equation \eqref{eq92} shows the 'A', 'B', and 'C' inputs of the DSP block using \eqref{eq8}. Each part is used as the input of the DSP block. The second row of the \eqref{eq8} is given to the 'A' input (25-bit) of the DSP block. The input variable is sent to the 'B' input (18-bit) of the DSP block. The 'C' input (48-bit) of the DSP block takes the third row of \eqref{eq8}. The 'A' input (25-bit) of the DSP block is multiplied with the 'B' input (18-bit) of the DSP block, and the result of the multiplication is accumulated with the 'C' input of the DSP block.

%\vspace{-1ex}
\begin{eqnarray}
\label{eq92}
\scalebox{0.9}{$ \displaystyle A = \sum_{i=2}^{k} \left( MW_{A_i} \ll (v + (i-1) \cdot (v + 3)) \right) + MW_{A_1} $} \\ 
\scalebox{0.9}{$ \displaystyle B = I \nonumber $} \\
\scalebox{0.9}{$ \displaystyle C = \sum_{i=1}^{k} \left(SEx_{A_i} \ll (i-1) \cdot (v + 3) \right) \nonumber $} 
\end{eqnarray}
%\vspace{1ex}

The post-processing part (Fig.~\ref{fig77}) takes the output of the DSP block as input and split it into three parts for 3 parameter multiplications. Final concatenation (with $I[n-1:0]$) and shift operations ($\ll s$) are employed for each part in parallel. Subsequently, S blocks perform sign conversion using the sign bits of parameters. 

Finally, accumulations are performed with the multiplication results to calculate the results of multiple MAC operations. LUTs are used for the final accumulation.

\section{Systolic Array Architecture}

\begin{figure}%[htbp]
\setlength\belowdisplayskip{0pt}
\centerline{\includegraphics[scale=0.35]{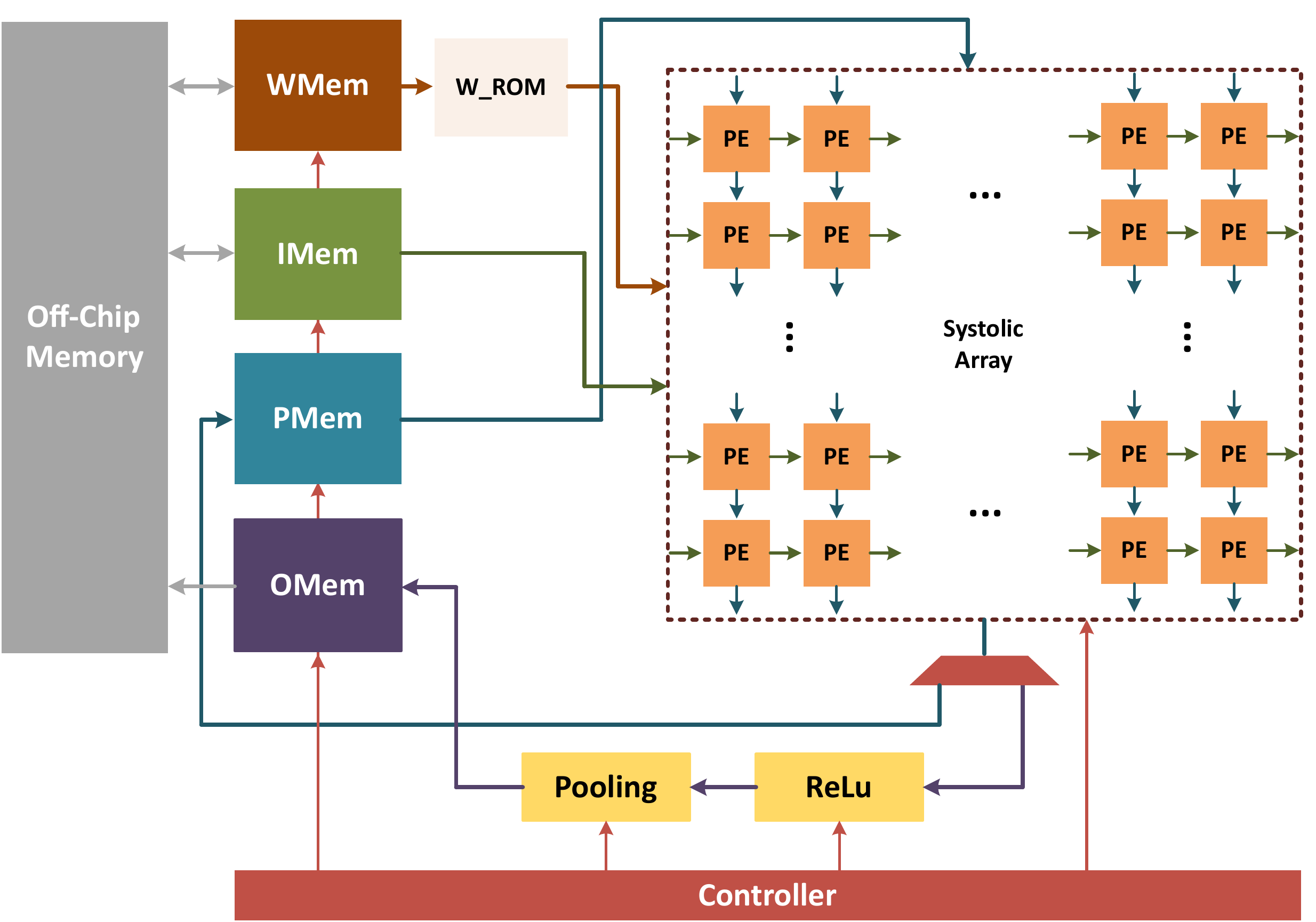}}
\caption{Systolic array architecture.}
\label{fig1}
%\vspace{-3.8ex}
\end{figure}

A top-level architecture of the prototype systolic array (SA) hardware is shown in Fig.~\ref{fig1}. Since the CNN models are employed as a use case, the designed systolic array architecture is customized for CNN inference. Four different memory blocks are implemented for on-chip data storage. In addition, on-chip ROM architecture is used as a dictionary to generate manipulated DSP inputs. 4, 6, and 8 bit signed fixed-point parameter multiplications are supported. Depending on the parameter bit length and the input variable bit length, each processing element (PE) is designed to execute 3, 4, or 6 parameter multiplications per DSP block.

Four AXI-mapped memory blocks, which have multiple BRAMs, are used for on-chip data storage. OMem (Fig.~\ref{fig1}) store output results before sending them off-chip memory. PMem stores partial sums to reuse it during convolution operations. IMem stores the input values (I) for convolution multiplications. Unlike the other 3 memory blocks, WMEM stores the address values for WROM.

Since the 'A' input value is independent of the input variable as shown in \eqref{eq92}, it is possible to calculate the 'A' values for parameter tuples once and store them on the on-chip ROM. Notice, the shift values, which are necessary for the calculation of the 'C' input, do not change with the input variable (I). Thus, these shift values are also calculated once and stored on the on-chip ROM. These shift values configure the hardware designed to generate the 'C' input of the DSP block. Storing all the required values for all parameter tuples brings the MBs of on-chip memory overhead, which is not tolerable considering the on-chip data storage capacity of the state-of-the-art FPGAs. A solution to this memory overhead problem has been achieved thanks to our parameter approximation technique. Since the developed parameter approximation technique constraints the $MW$s, it also reduces the number of possible parameter tuples. This makes the memory overhead problem manageable. As shown in Fig. \ref{fig12}, instead of storing the required values for ten different parameter tuples, the required values are stored for only two different parameter tuples thanks to our parameter approximation technique.

\begin{table*}
  \centering
  \renewcommand{\arraystretch}{1.2}
  \caption{Compression Rates (CONV Layers)}
  \begin{tabular}{*{8}{c}}
    \toprule
    (W,I) & \multicolumn{2}{c}{CNN Model} & DC \cite{d1} & H & WRC & WRC + H & P + WRC + H \\
    \midrule
    \multirow{2}{*}{(8,8)} & Alexnet & Conv & 9.09\% (11.0$\times$) & 14.65\% (6.8$\times$) & 66.6\% (1.5$\times$) & 10.80\% (9.3$\times$) & 8.96\% (11.2$\times$) \\
%    \cline{2-8}
    & VGG-16 & Conv & 7.28\% (13.7$\times$) & 14.18\% (7.0$\times$) & 66.6\% (1.5$\times$) & 10.17\% (9.8$\times$) & 8.49\% (11.8$\times$) \\
    \midrule 
    \multirow{2}{*}{(6,6)} & Alexnet & Conv & --- & 8.73\% (11.5$\times$) & 75.0\% (1.3$\times$) & 6.71\% (14.9$\times$) & 6.07\% (16.5$\times$) \\
%    \cline{2-8}
    & VGG-16 & Conv & --- & 8.10\% (12.3$\times$) & 75.0\% (1.3$\times$) & 6.10\% (16.4$\times$) & 5.64\% (17.7$\times$) \\
    \midrule  
    \multirow{2}{*}{(4,4)} & Alexnet & Conv & --- & 3.67\% (27.2$\times$) & 83.3\% (1.2$\times$) & 4.26\% (23.5$\times$) & 3.07\% (32.6$\times$) \\
%    \cline{2-8}
    & VGG-16 & Conv & --- & 3.29\% (30.4$\times$) & 83.3\% (1.2$\times$) & 3.77\% (26.5$\times$) & 2.97\% (33.6$\times$) \\
%    \hline \\ [-1ex]
  \bottomrule
    \multicolumn{8}{c}{(W,I): Bit Lengths of W and I, DC: Deep Compression, H: Huffman Coding, WRC: Parameter Representation Change, P: Pruning} 
  \end{tabular}
  \label{tab1}
\end{table*} 

Since the WROM stores, the 'A' input values, and the shift values required to generate the 'C' input values, all the information required to perform the parameter multiplication can be obtained from WROM. Consequently, there is no need to store actual parameters on the off-chip memory and the on-chip WMem. Instead, each parameter tuple can be represented as an index value of the WROM in the off-chip memory and on-chip WMEM. The index values stored in the off-chip memory and the on-chip WMem consist of the address of the WROM and the sign bits of the parameters in the parameter tuple. For example, a 16-bit address value is stored for each parameter tuple consisting of 8-bit fixed-point parameters. The most significant 13-bits are used to index the WROM, while the least significant 3-bit stores the sign bits of the 3 parameters. This enables 3x8-bit parameters to be represented with 16-bit in the off-chip memory and the on-chip WMem. The parameter representation change provides 33\% compression (Compression Rate: 66.6\%) of the parameters and reduces the access rate to the off-chip memory by a third without any hardware overhead. It is also possible to apply other compression techniques such as Huffman coding and pruning combined with the parameter representation change (WRC). Table \ref{tab1} shows the compression performance of the Alexnet and VGG-16 networks using parameter representation change, Huffman coding, and pruning. These results were also compared with the Deep Compression \cite{d1}. 

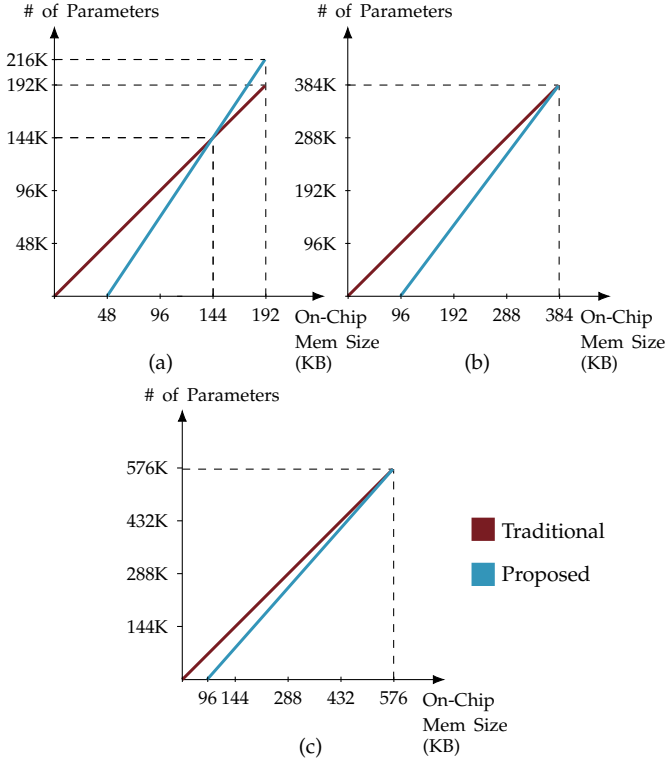
\begin{figure}[h]
\centering
\begin{tikzpicture}[PENode/.style={circle, draw, rectangle, thick, minimum size=8}]

\pgfmathtruncatemacro\zero{1}
\pgfmathtruncatemacro\four{5}

%
%
% GRAPH #1
%
%
\node[text width = 2cm] at (0.6,3.8) {\scriptsize\# of Parameters};
%\node[text width = 1.4cm] at (3.9,-0.7) {\footnotesize On-Chip Mem Size (KB)};

\node[text width = 0.5cm] at (1.5,-0.9) {\footnotesize (a)};

\node[text width = 1.4cm] at (3.9,-0.3) {\scriptsize On-Chip};
\node[text width = 1.4cm] at (3.9,-0.6) {\scriptsize Mem Size};
\node[text width = 1.4cm] at (3.9,-0.9) {\scriptsize (KB)};

\draw[arrows = {-Latex}] (0,0) -- (0,3.6) node[anchor=south west] {};
\draw[arrows = {-Latex}] (0,0) -- (3.6,0) node[anchor=north west] {};
\draw[dashed] (2.81,0) -- (2.81,3.15) node[anchor=north west] {};
\draw[dashed] (0,2.81) -- (2.8,2.81) node[anchor=north west] {};
\draw[dashed] (0,3.15) -- (2.8,3.15) node[anchor=north west] {};
\draw[dashed] (2.11,0) -- (2.11,2.1) node[anchor=north west] {};
\draw[dashed] (0,2.11) -- (2.1,2.11) node[anchor=north west] {};
\draw[line width=0.4mm, color=cayenne] (0,0) -- (2.8,2.8) node[anchor=north west] {};
\draw[line width=0.4mm, color=carrot!80] (0.7,0) -- (2.8,3.15) node[anchor=north west] {};

\foreach \x in {0,1,2,3, 4}
{
  \ifnum\x<\zero
    \draw (20*\x pt, 1pt) -- (20*\x pt, -1pt) node[anchor=north] {};
  \else
    \draw (20*\x pt, 1pt) -- (20*\x pt, -1pt) node[anchor=north] {\scriptsize \pgfmathparse{int(multiply(48,\x))} \pgfmathresult};
  \fi
}

\foreach \y in {0,1,2,3,4,5}
{
  \ifnum\y<\zero
    \draw (-1pt, 20*\y pt) -- (1pt, 20*\y pt) node[anchor=east] {};
  \else
    \ifnum\y<\four
      \draw (-1pt, 20*\y pt) -- (1pt, 20*\y pt) node[anchor=east] {\scriptsize \pgfmathparse{int(multiply(48,\y))} \pgfmathresult K};
    \else
      \draw (-1pt, 89.6 pt) -- (1pt, 89.6 pt) node[anchor=east] {\scriptsize 216K};
    \fi
  \fi
}

%
%
% GRAPH #2
%
%
\node[text width = 2cm] at (4.5,3.8) {\scriptsize \# of Parameters};
\node[text width = 1.4cm] at (7.7,-0.3) {\scriptsize On-Chip};
\node[text width = 1.4cm] at (7.7,-0.6) {\scriptsize Mem Size};
\node[text width = 1.4cm] at (7.7,-0.9) {\scriptsize (KB)};

\node[text width = 0.5cm] at (5.7,-0.9) {\footnotesize (b)};

%\node[text width = 2cm] at (4.2,-0.6) {Memory Size (KB)};

\draw[arrows = {-Latex}] (3.9,0) -- (3.9,3.6) node[anchor=south west] {};
\draw[arrows = {-Latex}] (3.9,0) -- (7.4,0) node[anchor=north west] {};

\draw[dashed] (6.71,0) -- (6.71,2.81) node[anchor=north west] {};
\draw[dashed] (3.9,2.81) -- (6.7,2.81) node[anchor=north west] {};

\draw[line width=0.4mm, color=cayenne] (3.9,0) -- (6.7,2.8) node[anchor=north west] {};
\draw[line width=0.4mm, color=carrot!80] (4.6,0) -- (6.7,2.8) node[anchor=north west] {};

\foreach \x in {0,1,2,3, 4}
{
  \ifnum\x<\zero
    \draw (111.1+20*\x pt, 1pt) -- (111.1+20*\x pt, -1pt) node[anchor=north] {};
  \else
    \draw (111.1+20*\x pt, 1pt) -- (111.1+20*\x pt, -1pt) node[anchor=north] {\scriptsize \pgfmathparse{int(multiply(96,\x))} \pgfmathresult};
  \fi
}

\foreach \y in {0,1,2,3,4}
{
  \ifnum\y<\zero
    \draw (112pt, 20*\y pt) -- (110pt, 20*\y pt) node[anchor=east] {};
  \else
    \draw (112pt, 20*\y pt) -- (110pt, 20*\y pt) node[anchor=east] {\scriptsize \pgfmathparse{int(multiply(96,\y))} \pgfmathresult K};
  \fi
}

%
%
% GRAPH #3
%
%
\node[text width = 2cm] at (2.2,-1.3) {\scriptsize \# of Parameters};
\node[text width = 1.4cm] at (5.6,-5.4) {\scriptsize On-Chip};
\node[text width = 1.4cm] at (5.6,-5.7) {\scriptsize Mem Size};
\node[text width = 1.4cm] at (5.6,-6.0) {\scriptsize (KB)};

%\node[text width = 2cm] at (4.2,-0.6) {Memory Size (KB)};

\node[text width = 0.5cm] at (3.5,-6.0) {\footnotesize (c)};

\draw[arrows = {-Latex}] (1.7,-5.1) -- (1.7,-1.5) node[anchor=south west] {};
\draw[arrows = {-Latex}] (1.7,-5.1) -- (5.2,-5.1) node[anchor=north west] {};

\draw[dashed] (4.51,-5.1) -- (4.51,-2.3) node[anchor=north west] {};
\draw[dashed] (1.7,-2.3) -- (4.5,-2.3) node[anchor=north west] {};

\draw[dashed] (0,3.15) -- (2.8,3.15) node[anchor=north west] {};

\draw[dashed] (2.11,0) -- (2.11,2.1) node[anchor=north west] {};
\draw[dashed] (0,2.11) -- (2.1,2.11) node[anchor=north west] {};

\draw[line width=0.4mm, color=cayenne] (1.7,-5.1) -- (4.5,-2.3) node[anchor=north west] {};
\draw[line width=0.4mm, color=carrot!80] (2.03,-5.1) -- (4.5,-2.3) node[anchor=north west] {};

\foreach \x in {0,1,2,3, 4,5}
{
  \ifnum\x<\zero
    \draw (48.4+20*\x pt, -144pt) -- (48.4+20*\x pt, -146pt) node[anchor=north] {};
  \else
    \ifnum\x<\four
      \draw (48.4+20*\x pt, -144pt) -- (48.4+20*\x pt, -146pt) node[anchor=north] {\scriptsize \pgfmathparse{int(multiply(144,\x))} \pgfmathresult};
    \else
      \draw (58pt, -144pt) -- (58pt, -146pt) node[anchor=north] {\scriptsize 96};
    \fi
  \fi
}

\foreach \y in {0,1,2,3,4}
{
  \ifnum\y<\zero
    \draw (48.4pt, 20*\y pt) -- (46.4pt, 20*\y pt) node[anchor=east] {};
  \else
    \draw (48.4pt, -145pt + 20*\y pt) -- (46.4pt, -145pt + 20*\y pt) node[anchor=east] {\scriptsize \pgfmathparse{int(multiply(144,\y))} \pgfmathresult K};
  \fi
}

\node[PENode, color=cayenne, fill=cayenne] (36) at (5.7, -3.1) {};
\node[PENode, color=carrot!80, fill=carrot!80] (36) at (5.7, -3.7) {};

\node[text width = 0.7cm] at (6.3,-3.12) {\footnotesize Traditional};
\node[text width = 0.7cm] at (6.3,-3.72) {\footnotesize Proposed};

\end{tikzpicture}
\caption{On-chip memory size analysis, (a) 8-bit parameters, (b) 6-bit parameters, (c) 4-bit parameters.}
\label{fig13}
%\vspace{-1.5ex}
\end{figure} 
As shown in Table \ref{tab1}, our technique combined with Huffman coding and pruning produces comparable results with the Deep Compression. Only WRC was implemented in the SA hardware. Compression results obtained using the Huffman coding and pruning were reported for analysis and comparison. 

Even though the approximation technique reduces the WROM size, it is still an overhead for the hardware implementation. However, the size of the WMem implemented in this hardware is less than the traditional implementations because parameters are represented on the WMem by fewer bits due to parameter representation change. This may compensate for the overhead caused by WROM. As shown in Fig.~\ref{fig13}, the number of parameters stored in on-chip memory with our hardware is higher than the traditional hardware implementations in case of the on-chip memory size is higher than a certain value. This analysis shows that using WROM may provide advantage instead of being overhead for hardware implementation. The on-chip memory sizes are given in Fig.~\ref{fig13} for traditional hardware implementations and the proposed hardware in this paper (for 4, 6, and 8 bit parameters). The initial points for our implementation in Fig.~\ref{fig13} shows the overhead caused by the WROM. After these initial overhead values, WMEM stores the parameters.

\begin{figure}[t]
\centering
\centerline{\includegraphics[scale=0.55]{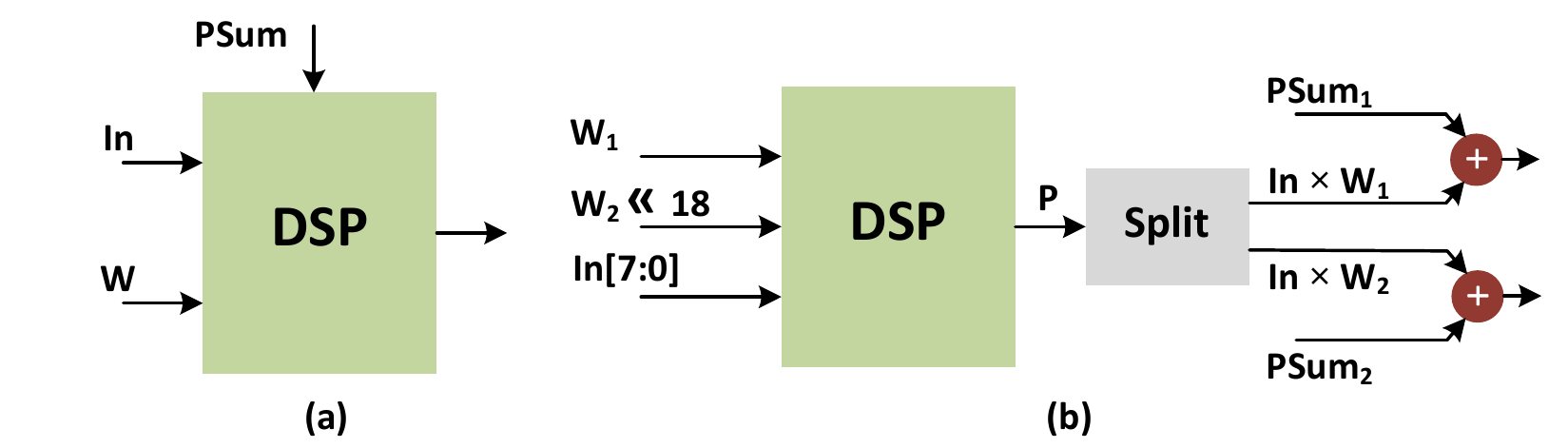}}
\caption{PE architecture (a) One MAC per DSP, (b) Two multiplications per DSP.}
\label{fig4}
%\vspace{-2ex}
\end{figure}

Fig.~\ref{fig4}(a) and (b) show two different PE architectures. Fig.~\ref{fig4}(a) shows a traditional PE architecture, which executes a 1 MAC operation per DSP block. The PE architecture, which is shown in Fig.~\ref{fig4}(b), can execute 2 8-bit parameter multiplications on a single DSP block, while it uses LUTs and FFs for the accumulation execution. PE architecture shown in Fig.~\ref{fig4}(b) uses the method presented in \cite{c2} to execute 2 multiplications per DSP block. The prototype systolic array hardware is also implemented using the PE architecture given in Fig.~\ref{fig4}(a) and (b) to make a comparison with the PE architecture given in Fig.~\ref{fig77}. 

Weight stationary (WS) dataflow is used for the prototype SA architecture. Weights are initially loaded to PEs, and the inputs and PSums flow between the PEs. Since each input is multiplied with different CNN weights for different kernels in each channel, multiple CNN weights can be multiplied with one input. WS dataflow allows the reuse of the CNN weights in the PEs during convolution operations. The goal of using the WS dataflow is to reduce power consumption by reducing the switching on the parameter decompression hardware.

\section{Implementation Results}

The prototype systolic array (SA) architecture is synthesized, placed, and routed for a Xilinx Zynq-7000 ZC706 FPGA using Xilinx Vivado 18.3. Three different parameter bit lengths (4, 6, and 8) are supported in the implemented SA architecture. The implementation results are given in Table \ref{tab21} for the 12$\times$12 SA. Since the number of parameters that can be multiplied on a single DSP block changes based on the input variable bit length, the number of DSP blocks reported in Table \ref{tab21} is different for the different bit lengths. The number of DSP blocks given in Table \ref{tab21} indicates that as the input variable bit-length decreases, our technique can perform much more multiplications on a single DSP block. The number of LUTs required for the accumulation and the overhead caused by the parameter decompression are also reported in Table \ref{tab21}. 

\begin{table}
  \renewcommand{\arraystretch}{1.2}
  \caption{Implementation Results (12 $\times$ 12 PEs)}
  \label{tab21}
  \centering
  \begin{tabular}{*{5}{c}}
    \toprule
    & & 4-bit & 6-bit & 8-bit \\ 
    & &(6M/DSP)&(4M/DSP)&(3M/DSP)\\ 
    \midrule
    \multicolumn{2}{c}{\multirow{2}{*}{FPGA}} & Xilinx & Xilinx & Xilinx  \\
    & & Zynq & Zynq & Zynq  \\
    \midrule
    \multirow{3}{*}{LUT}& P Decomp. &432 & 972 & 1680 \\
    & Post-P. &576 & 2016 & 3769 \\
    & Accum. &1152 & 1728 & 2160 \\
    \midrule
    \multicolumn{2}{c}{DFF}&5732 & 7667 & 9244 \\
    \midrule
    \multicolumn{2}{c}{DSP}&24 & 36 & 48  \\
    \midrule
    \multicolumn{2}{c}{BRAM}&54 & 68.5 & 69  \\
    \midrule
    \multicolumn{2}{c}{Freq.}&250 & 250 & 250  \\
  \bottomrule
\end{tabular}
\end{table}

\begin{table}
  \renewcommand{\arraystretch}{1.2}
  \caption{Hardware Comparison (12 $\times$ 12 PEs)}
  \label{tab2}
  \centering
  \begin{tabular}{*{7}{c}}
    \toprule
    Bit Length & Impl. & LUT & DFF & DSP & BRAM & Freq.\\ 
    \midrule
    \multirow{2}{*}{4}&1M& 235 & 10167 & 144 & 48 & 270\\
    &MP& 2356 & 5732 & 24 & 54 & 250\\
    \midrule
    \multirow{2}{*}{6}&1M& 382 & 11189 & 144 & 69.5 & 256\\
    &MP& 5459 & 7667 & 36 & 68.5 & 250\\
    \midrule
    \multirow{3}{*}{8}&1M& 475 & 11973 & 144 & 92 & 250\\
    &2M& 2773 & 8343 & 72 & 92 & 250\\
    &MP& 8217 & 9244 & 48 & 69 & 250\\
  \bottomrule
\end{tabular}
\end{table}

\begin{table}[!t]
  \renewcommand{\arraystretch}{1.2}
  \caption{Hardware Comparison with Xilinx DPU (256 PEs)}
  \label{tab3}
  \centering
  \begin{tabular}{*{6}{c}}
    \toprule
    \multirow{2}{*}{Impl.}  & \multirow{2}{*}{LUT} & \multirow{2}{*}{DFF} & \multirow{2}{*}{DSP} & \multirow{2}{*}{BRAM} & Peak Perf.\\
    & & & & & (GOPs) \\
    \midrule
    DPUH& 20055 & 28849 & 98 & 69.5 & 102\\
    DPUL& 21171 & 33572 & 66 & 69.5 & 102\\
    MP& 11562 & 13882 & 88 & 76 & 128\\
  \bottomrule
\end{tabular}
\end{table} 

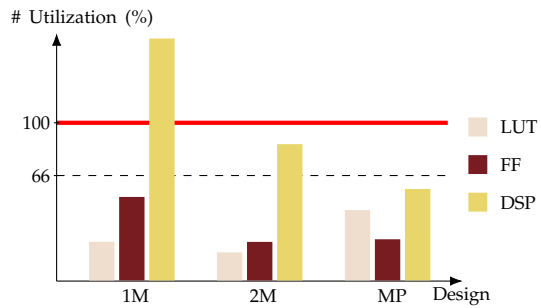
\begin{figure}[t]
\centering

\pgfdeclarelayer{background layer}
\pgfdeclarelayer{background layer 2}
\pgfsetlayers{background layer, background layer 2, main}

\begin{tikzpicture}[PENode/.style={circle, draw, rectangle, thick, minimum size=5}]

\pgfmathtruncatemacro\zero{1}
\pgfmathtruncatemacro\four{5}

\node[text width = 2cm] at (-2.2,3.2) {\scriptsize\# Utilization (\%)};

\node[text width = 1.4cm] at (3.1,-0.5) {\scriptsize Design};

\draw[arrows = {-Latex}] (-2.6,-0.3) -- (-2.6,3) node[anchor=south west] {};
\draw[arrows = {-Latex}] (-2.6,-0.3) -- (2.8,-0.3) node[anchor=north west] {};

\node[rectangle, draw, ultra thick, color = ash, fill=ash, minimum size = 8, minimum height = 13] at (-2,-0.035) {};
\node[rectangle, draw, ultra thick, color = cayenne, fill=cayenne, minimum size = 8, minimum height = 30] at (-1.6,0.264) {};
\node[rectangle, draw, ultra thick, color = yllow, fill=yllow, minimum size = 8, minimum height = 90] at (-1.2,1.318) {};
\node[text width = 1.4cm] at (-1.03,-0.5) {\scriptsize 1M};

\node[rectangle, draw, ultra thick, color = ash, fill=ash, minimum size = 8, minimum height = 9] at (-0.3,-0.105) {};
\node[rectangle, draw, ultra thick, color = cayenne, fill=cayenne, minimum size = 8, minimum height = 13] at (0.1,-0.035) {};
\node[rectangle, draw, ultra thick, color = yllow, fill=yllow, minimum size = 8, minimum height = 50] at (0.5,0.615) {};
\node[text width = 1.4cm] at (0.67,-0.5) {\scriptsize 2M};

\node[rectangle, draw, ultra thick, color = ash, fill=ash, minimum size = 8, minimum height = 25] at (1.4,0.177) {};
\node[rectangle, draw, ultra thick, color = cayenne, fill=cayenne, minimum size = 8, minimum height = 14] at (1.8,-0.018) {};
\node[rectangle, draw, ultra thick, color = yllow, fill=yllow, minimum size = 8, minimum height = 33] at (2.2,0.317) {};
\node[text width = 1.4cm] at (2.37,-0.5) {\scriptsize MP};

\begin{pgfonlayer}{background layer 2}

\draw[dashed, below] (-2.6,1.104) -- (2.6,1.104) node[anchor=north west] {};
\draw[below, color = red, ultra thick] (-2.6,1.806) -- (2.6,1.806) node[anchor=north west] {};

\end{pgfonlayer}{background layer 2}

\draw (-75pt, 1.104) -- (-73pt, 1.104) node[anchor=east] {\scriptsize 66};
\draw (-75pt, 1.806) -- (-73pt, 1.806) node[anchor=east] {\scriptsize 100};

\node[PENode, color=ash, fill=ash] (40) at (3, 1.75) {};
\node[PENode, color=cayenne, fill=cayenne] (40) at (3, 1.25) {};
\node[PENode, color=yllow, fill=yllow] (40) at (3, 0.75) {};

\node[text width = 0.7cm] at (3.65,1.75) {\scriptsize LUT};
\node[text width = 0.7cm] at (3.65,1.25) {\scriptsize FF};
\node[text width = 0.7cm] at (3.65,0.75) {\scriptsize DSP};

\end{tikzpicture}
\caption{FPGA resource utilization analysis (Zybo Z7-10)}
\label{fig330}
%\vspace{-1.5ex}
\end{figure}

Three different versions of the SA prototype are implemented using different PE architectures; (a) One MAC per DSP block (1M), (b) two parameter multiplications per DSP block (2M) and (c) multiple parameter multiplications per DSP block (SDMM) including the novel approximation technique and the parameter representation change (multiplication packing - MP). Table \ref{tab2} shows the implementation results for the three different PE architectures. The implementation results for 2M are given only for 8-bit parameters because 2M can supports only a multiplication with 8-bit parameters. As shown in Table \ref{tab2}, the SA hardware including the parameter approximation, multiplication packing (SDMM), and the fine-tuning techniques reduced the number of DSP blocks used in the baseline FPGA implementation (1M) by 66.6\%, 75\%, and 83.3\% for 8, 6, and 4-bit implementations, respectively. We also synthesized 1M, 2M, and MP implementations of 8-bit parameters on Xilinx Zybo Z7-10 to analyze the efficiency of our design on low-cost FPGAs. As shown in Fig.~\ref{fig330}, our MP implementation only uses 60\% of available DSP blocks while 1M could not fit into the Zybo Z7-10 FPGA board.

\begin{figure}[t]
\centering

\pgfdeclarelayer{background layer}
\pgfdeclarelayer{background layer 2}
\pgfsetlayers{background layer, background layer 2, main}

\begin{tikzpicture}[PENode/.style={circle, draw, rectangle, thick, minimum size=5}]

\pgfmathtruncatemacro\zero{1}
\pgfmathtruncatemacro\four{5}

\node[text width = 2cm] at (-2.2,3.2) {\scriptsize\# Power (mW)};

\node[text width = 1.4cm] at (3.0,-0.5) {\scriptsize Parameter};
\node[text width = 1.4cm] at (3.0,-0.8) {\scriptsize Bit-Length};

\draw[arrows = {-Latex}] (-2.6,-0.3) -- (-2.6,3) node[anchor=south west] {};
\draw[arrows = {-Latex}] (-2.6,-0.3) -- (2.8,-0.3) node[anchor=north west] {};

\node[rectangle, draw, ultra thick, color = ash, fill=ash, minimum size = 12, minimum height = 76.5] at (-2,1.08) {};
\node[rectangle, draw, ultra thick, color = cayenne, fill=cayenne, minimum size = 12, minimum height = 26.5] at (-1.4,0.2) {};
\node[text width = 1.4cm] at (-1.2,-0.5) {\scriptsize 4-bit};
\node[text width = 1.4cm] at (-1.4,-0.75) {\scriptsize (6 MAC)};

\node[rectangle, draw, ultra thick, color = ash, fill=ash, minimum size = 12, minimum height = 62] at (-0.3,0.82) {};
\node[rectangle, draw, ultra thick, color = cayenne, fill=cayenne, minimum size = 12, minimum height = 26.5] at (0.3,0.2) {};
\node[text width = 1.4cm] at (0.5,-0.5) {\scriptsize 6-bit};
\node[text width = 1.4cm] at (0.3,-0.75) {\scriptsize (4 MAC)};

\node[rectangle, draw, ultra thick, color = ash, fill=ash, minimum size = 12, minimum height = 48] at (1.4,0.58) {};
\node[rectangle, draw, ultra thick, color = cayenne, fill=cayenne, minimum size = 12, minimum height = 30] at (2.0,0.26) {};
\node[text width = 1.4cm] at (2.2,-0.5) {\scriptsize 8-bit};
\node[text width = 1.4cm] at (2.0,-0.75) {\scriptsize (3 MAC)};

\begin{pgfonlayer}{background layer 2}
\draw[dashed, below] (-2.6,0.4) -- (2.6,0.4) node[anchor=north west] {};
\draw[dashed, below] (-2.6,1.104) -- (2.6,1.104) node[anchor=north west] {};
\draw[dashed, below] (-2.6,1.806) -- (2.6,1.806) node[anchor=north west] {};
\draw[dashed, below] (-2.6,2.510) -- (2.6,2.510) node[anchor=north west] {};
%\draw[dashed, below] (-2.6,2.811) -- (2.6,2.811) node[anchor=north west] {};
\end{pgfonlayer}{background layer 2}

\draw (-75pt, 0.4) -- (-73pt, 0.4) node[anchor=east] {\scriptsize 10};
\draw (-75pt, 1.104) -- (-73pt, 1.104) node[anchor=east] {\scriptsize 20};
\draw (-75pt, 1.806) -- (-73pt, 1.806) node[anchor=east] {\scriptsize 30};
\draw (-75pt, 2.510) -- (-73pt, 2.510) node[anchor=east] {\scriptsize 40};

\node[PENode, color=ash, fill=ash] (40) at (3, 1.75) {};
\node[PENode, color=cayenne, fill=cayenne] (40) at (3, 1.25) {};

\node[text width = 0.7cm] at (3.65,1.75) {\scriptsize 1M};
\node[text width = 0.7cm] at (3.65,1.25) {\scriptsize MP};

\end{tikzpicture}
\caption{Power consumption comparison}
\label{fig333}
%\vspace{-1.5ex}
\end{figure}
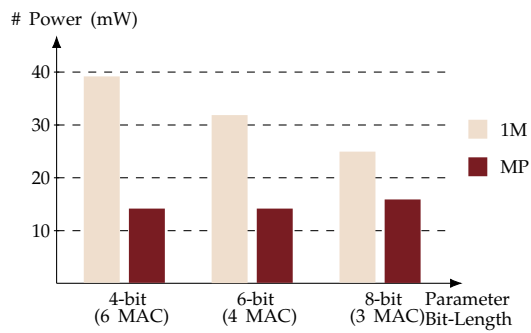

The SA hardware using our technique is also compared with the Xilinx DPU as shown in Table \ref{tab3}. the DPUH and DPUL refer to a high and low DSP usage configurations of the DPU. As shown in Table \ref{tab3}, the MP uses a fewer number of DSP blocks than the Xilinx DPUH. Also, the DPUH uses more LUTs and FFs compared to the MP. In the DPUH, MAC operations are shared between the LUTs/FFs and the DSP blocks during hardware synthesis. Also, in the DPUH, one DSP block can execute one MAC operation only. On the other hand, only multiplication operations are executed on the DSP blocks in the DPUL. Similar to the DPUH, the DPUL shared the multiplication operations between the LUTs/FFs and the DSP blocks. Accumulation operations are executed only on the LUTs/FFs. As a result, the DPUL uses a fewer number of DSP blocks than the MP, but it needs more than twice the LUTs as the MP. 

We also compare the power consumption of the 1M and MP for 4, 6, and 8-bit signed parameters. The MP can implement 6, 4, and 3 multiplications on a single DSP block for 4, 6, and 8-bit signed parameters, respectively. Hence, we implemented 6, 4, and 3 MAC calculation blocks to estimate the power consumption of both the 1M and MP for 4, 6, and 8-bit signed parameters, respectively. The Xilinx Vivado tool is used to estimate the power consumption of the 1M and MP for 4, 6, and 8-bit signed parameters. All switching activities are stored in SAIF files during post-implementation timing simulation. The Xilinx Vivado tool read these SAIF files and estimates the power consumption of the 1M and MP for 4, 6, and 8-bit signed parameters. As shown in Fig.~\ref{fig333}, our MP implementation reduces the power consumption of the 1M by 64.1\%, 54.8\%, and 36\% for 4, 6, and 8-bit signed parameters, respectively.

\section{Conclusions}
In this paper, an efficient parameter approximation technique was introduced to pack multiplication operations and perform multiple parameter multiplications on a single DSP block (SDMM). This was achieved by redeployment of the accumulator component of the traditional MAC operation and insertion of a parallel addition component. This method performs a mathematical manipulation on low bit-length fixed-point parameters and reduces their bit-lengths using an efficient approximation. Also, it employs a fine-tuning step to guarantee that the number of parameter multiplications per DSP block is fixed. Accuracy of the newly presented parameter approximation technique was measured using the Alexnet and VGG-16 networks and the Tiny ImageNet dataset for different bit lengths. This optimization leads to minor increases or decreases in the accuracy of reduced bit length implementation of the Alexnet and VGG-16. A prototype systolic array architecture, which used a processing element that can execute multiple parameter multiplications per DSP block, was implemented on a Xilinx Zynq-7000 ZC706 FPGA. This prototype reduced the number of DSP blocks used in the baseline FPGA implementation by 66.6\%, 75\%, and 83.3\% for the 8, 6, and 4-bit input variables. Additionally, 33\% compression was achieved by changing the parameter representation.

\ifCLASSOPTIONcompsoc
  \section*{Acknowledgments}
\else
  \section*{Acknowledgment}
\fi

This work was supported by the EU ECSEL Joint Undertaking, which funded the NewControl project under the grant agreement No. 826653-2.

\ifCLASSOPTIONcaptionsoff
  \newpage
\fi

\end{document}